\documentclass[a4paper, 10pt]{amsart}
\usepackage[utf8]{inputenc}
\usepackage{fullpage,
    todonotes,
    amsmath,
    amsfonts,
    amsthm,
    isomath,
    booktabs,
    siunitx,
    algorithm,
    algorithmic,
    pgfplots, %
    rotating, %
    subcaption, %
    mathtools} %
\usepackage[hidelinks]{hyperref}

\usepackage[foot]{amsaddr}
\usepgfplotslibrary{colorbrewer}
\usepgflibrary{plotmarks}

\graphicspath{{./figures}}

\theoremstyle{remark}
\newtheorem{remark}{Remark}

\renewcommand*{\vec}{\vectorsym}
\newcommand{\ten}{\mathsf}

\DeclareMathOperator{\diver}{div}
\DeclareMathOperator{\grad}{grad}
\DeclareMathOperator{\Diver}{Div}
\DeclareMathOperator{\Grad}{Grad}
\DeclareMathOperator{\cof}{Cof}

\DeclareMathOperator{\lin}{Lin}

\DeclareMathOperator{\tr}{tr}

\newcommand{\cauchy}{\ten{T}}

\newcommand{\eshelby}{\ten{\Sigma}}

\newcommand{\cur}[1]{\widehat{#1}} 
\newcommand{\tvchi}{\cur{\vec \chi}}
\newcommand{\tu}{\cur{\vec u}}
\newcommand{\tF}{\cur{\ten F}}
\newcommand{\tPsi}{\cur{\Psi}}

\renewcommand{\d}{\mathrm{d}}
\newcommand{\dX}{\,\d X}
\newcommand{\dx}{\,\d x}

\newcommand{\TensAct}{T_{\mathrm{a}}}
\newcommand{\fiberRefF}{\vec e_f}
\newcommand{\fiberRefS}{\vec e_s} 
\newcommand{\fiberRefN}{\vec e_n}
\newcommand{\fiberRef}[1]{\vec e_{#1}}
\newcommand{\ratioN}{\alpha_n}
\newcommand{\BCKn}{K_{\mathrm{n}}}
\newcommand{\BCKt}{K_{\mathrm{t}}}

\newcommand{\numCavities}{N_{\mathrm{cav}}}
\newcommand{\bdryEndoREF}[1]{\Gamma^{\mathrm{endo,#1}}_0}
\newcommand{\bdryEndoACT}[1]{\Gamma^{\mathrm{endo,#1}}}
\newcommand{\bdryEpiREF}{\Gamma^{\mathrm{epi}}_0}
\newcommand{\bdryEpiACT}{\Gamma^{\mathrm{epi}}}
\newcommand{\bdryDREF}{\Gamma^{\mathrm{D}}_0}
\newcommand{\bdryDACT}{\Gamma^{\mathrm{D}}}

\usepackage{listings}
\usepackage{xcolor}
\definecolor{codegreen}{rgb}{0,0.6,0}
\definecolor{codegray}{rgb}{0.5,0.5,0.5}
\definecolor{codepurple}{rgb}{0.58,0,0.82}
\definecolor{backcolour}{rgb}{0.95,0.95,0.92}
\lstdefinestyle{mystyle}{
  backgroundcolor=\color{backcolour}, commentstyle=\color{codegreen},
  keywordstyle=\color{magenta},
  numberstyle=\tiny\color{codegray},
  stringstyle=\color{codepurple},
  basicstyle=\ttfamily\footnotesize,
  breakatwhitespace=false,         
  captionpos=b,                    
  keepspaces=true,                 
  numbers=none,                    
  numbersep=5pt,                  
  showspaces=false,                
  showstringspaces=false,
  showtabs=false,                  
  tabsize=2,
  framerule=1.5pt,
  rulecolor=\color{red!60!black}
}
\DeclareSIUnit\mmHg{mmHg}

\begin{document}

\title{Reconstructing relaxed configurations in elastic bodies: Mathematical formulations and numerical methods for cardiac modeling}

\author{N. A. Barnafi$^1$}
\email{nbarnafi@cmm.uchile.cl}
\author{F. Regazzoni$^2$}
\email{francesco.regazzoni@polimi.it}
\thanks{Corresponding author: F. Regazzoni.}
\author{D. Riccobelli$^2$}
\email{davide.riccobelli@polimi.it}
\address[1]{Centro de Modelamiento Matemático, Universidad de Chile,
    Av. Beauchef 851, Santiago, Chile}
\address[2]{MOX -- Dipartimento di Matematica, Politecnico di Milano,
    Piazza Leonardo da Vinci 32, Milano, 20133, Italy}

\begin{abstract}
    Modeling the behavior of biological tissues and organs often necessitates the knowledge of their shape in the absence of external loads.
    However, when their geometry is acquired \emph{in-vivo} through imaging techniques, bodies are typically subject to mechanical deformation due to the presence of external forces, and the load-free configuration needs to be reconstructed.
    This paper addresses this crucial and frequently overlooked topic, known as the inverse elasticity problem (IEP), by delving into both theoretical and numerical aspects, with a particular focus on cardiac mechanics.
    In this work, we extend Shield's seminal work to determine the structure of the IEP with arbitrary material inhomogeneities and in the presence of both body and active forces.
    These aspects are fundamental in computational cardiology, and we show that they may break the variational structure of the inverse problem.
    In addition, we show that the inverse problem might have no solution even in the presence of constant Neumann boundary conditions and a polyconvex strain energy functional.
    We then present the results of extensive numerical tests to validate our theoretical framework, and to characterize the computational challenges associated with a direct numerical approximation of the IEP.
    Specifically, we show that this framework outperforms existing approaches both in terms of robustness and optimality, such as Sellier's iterative procedure, even when the latter is improved with acceleration techniques.
    A notable discovery is that multigrid preconditioners are, in contrast to standard elasticity, not efficient, where a one-level additive Schwarz and generalized Dryja-Smith-Widlund provide a much more reliable alternative.
    Finally, we successfully address the IEP for a full-heart geometry, demonstrating that the IEP formulation can compute the stress-free configuration in real-life scenarios where Sellier's algorithm proves inadequate.
\end{abstract}

\maketitle

\section{Introduction}

The objects we observe are rarely free from external mechanical stresses. For example, all bodies around us are subject to gravity. While such a force usually induces small displacements in stiffer materials, it can lead to large deformations in soft matter~\cite{Mora_2014,Riccobelli_2017,mora2019shape}. Furthermore, in biomedical applications, the shape of organs and tissues observed through medical imaging techniques are affected by the presence of mechanical forces that can significantly deform them. An important example in this respect is the heart: the presence of the ribcage and surrounding organs, as well as the blood pressure in the chambers, produce large deformations and it is not possible to observe its relaxed shape \emph{in-vivo}. In fact, directly observing the configuration of an elastic body in the absence of external forces is far from being a trivial task.

In nonlinear elasticity, the task of reconstructing the relaxed configuration of a body subject to mechanical loads, hereafter referred to as the \emph{inverse elasticity problem} (IEP), is a long-standing and largely overlooked problem, which is briefly cited in the Truesdell and Noll book as the \emph{free shape problem}~\cite{truesdell2013non}.
The problem has received little attention from the continuum mechanics community: it has been originally addressed by Shield~\cite{Shield_1967} for homogeneous bodies in the absence of body forces, and has been extended by Merodio and Ogden~\cite{Merodio_2006} to take into account body forces. Up to our knowledge, Shield's theory has never been extended to the inhomogeneous case, despite it being fundamental in several application areas, including computational cardiology, since the fiber direction changes within the myocardium. Shield's theory has been exploited as a tool to identify analytical solutions in non-linear elasticity~\cite{Carroll_2005,murphy2003inverse,horgan2004invariance,horgan2005plane,carroll2005compressible}, but the structure of the IEP as a boundary value problem remains largely unexplored.

The IEP has received some more attention in the scientific computing community~\cite{mazier2022inverse}, where it is known as \emph{inverse design problem}~\cite{govindjee1996computational} or \emph{prestress problem}~\cite{gee2009prestressing}.
Its role in the specific case of cardiac modeling, and in biomechanics in general, is pivotal, as a reliable identification of the relaxed configuration is fundamental to correctly describe the stress distribution in soft tissues~\cite{peirlinck2018modular,marx2022robust,regazzoni2022emcirculation}. We highlight also an application to poromechanics in \cite{patte2022quasi}. A possible solution approach, based on a fixed-point algorithm, was proposed by Sellier~\cite{sellier2011iterative} and allows for solving the inverse problem by leveraging only a solver for the direct problem. This approach is particularly attractive, as it allows to re-use existing software. However, when applied to real-life problems such as four-chamber cardiac geometries, it often presents convergence issues. To mitigate them, Sellier's method has been improved through adaptive continuation methods~\cite{regazzoni2022emcirculation} and acceleration techniques~\cite{rausch2017augmented,marx2022robust}. We highlight that the Sellier's method is not only relevant for cardiac simulations, and it has indeed also been used for modeling the eyes~\cite{montanino2020recovery}, aorta~\cite{gee2009prestressing}, and brain~\cite{Morin_2015}. An interesting problem, that we do not address in this work, is that of modeling the residual stresses in the tissue. See \cite{genet2015heterogeneous} for more details on this topic.

In this work, we study the IEP, with a special focus on the context of cardiac modeling. Our scope is twofold: on one hand, we study the mathematical structure of the IEP, extending Shield's theory to the case of inhomogeneous bodies subject to active forces. %
On the other hand, we thoroughly characterize this problem numerically for increasing levels of complexity and compare a direct numerical approximation of the latter with the Sellier method in terms of robustness with respect to external loads and its optimality.

The paper is organized as follows. In Section~\ref{section:problem-description}, we review some basic facts of non-linear elasticity and we derive the IEP, together with some remarks on the mathematical structure and some elementary examples. %
In Section~\ref{section:numerical-approximation}, we derive the weak formulation for both the direct and the inverse elasticity problems. In Section~\ref{section:physical-ref-conf}, we show with simple examples the mechanisms through which the IEP problem can give rise to self-intersections. In Section~\ref{section:algorithms} we describe all the algorithms that we consider for this study, which are (a) the Sellier method, (b) the Aitken accelerated Sellier method, (c) the Anderson accelerated Sellier method, and (d) the direct numerical approximation of the IEP. In Section~\ref{section:numerical-tests} we provide several numerical tests with the scope of (a) validating our theoretical claims, (b) characterizing the computational burden of IEP, and (c) testing the methods in realistic cardiac contexts. We conclude our work in Section~\ref{section:conclusions}.

\section{Problem description}\label{section:problem-description}
In this section we describe both the direct elasticity problem (DEP) and the inverse elasticity problem (IEP). For the latter, we show how it can be re-cast in terms of the Eshelby tensor, which will provide a way to guarantee the existence of solutions of the IEP under some special conditions.

\subsection{The direct problem of non-linear elasticity}
We assume that a body occupies a given region $\Omega_0$ of the three dimensional Euclidean space $\mathbb{E}^3$.
Let $\Omega\subset\mathbb{E}^3$ be the current configuration of the body, which is given by a deformation field $\vec{\chi}$ such that $\Omega=\vec{\chi}(\Omega_0)$. Specifically, the current position of the generic material point $\vec X\in\Omega_0$ is denoted by $\vec x$, i.e. $\vec{x} = \vec{\chi}(\vec X)$.
The displacement field is thus defined as  $\vec{u}(\vec{X})\coloneqq\vec{\chi}(\vec{X})-\vec{X}$.
We denote by $\Grad$ and $\grad$ the gradient operators with respect to $\vec{X}$ and $\vec{x}$, respectively. Similarly, we denote by $\Diver$ and $\diver$ the corresponding divergence operators.

We introduce the deformation gradient $\ten{F}\coloneqq\Grad\vec{\chi}=\ten I + \Grad \vec u$, together with the local volume change described by $J\coloneqq\det\ten F$.
Let $\ten P$ be the Piola-Kirchhoff stress tensor, then under the assumption of quasi-static deformations, the balance of the linear momentum reads
\begin{equation}
    \label{eq:balance_linear}
    \Diver\ten P + \vec{B}=\vec 0, \quad \text{in $\Omega_0$},
\end{equation}
where $\vec{B}$ is the density of body forces in the reference configuration. Such a balance equation can be also cast in current configuration by means of the Cauchy stress tensor
\begin{equation}
    \label{eq:Cauchy}
    \ten T = J^{-1} \ten P\ten F^{T}.
\end{equation}
More specifically, we get
\begin{equation}
    \label{eq:balance_linear_cauchy}
    \diver\ten T + \vec{b} = \vec{0},
\end{equation}
where $\vec{b}$ is the density of body force in the current configuration. The material and the spatial densities of force $\vec B$ and $\vec b$ are related by $\vec{B}=J\vec{b}$. For illustrative purposes, in this section, we assume that the boundary $\partial\Omega_0$ is composed of two distinct subsets, $\Gamma^{\mathrm{D}}_0$ and $\Gamma^{\mathrm{N}}_0$, such that on $\Gamma^{\mathrm{D}}_0$ we prescribe the displacement field $\vec u_{\mathrm{D}}$, and on $\Gamma_0^{\mathrm{N}}$ we assume that
\begin{equation}
    \label{eq:BC_lagr}
    \ten P\vec{N} = \vec{t}_0\qquad\text{on }\Gamma^{\mathrm{N}}_0,
\end{equation}
where the traction load $\vec{t}_0$ is a (known) vector field over $\Gamma^{\mathrm{N}}_0$. The Eulerian counterpart of \eqref{eq:BC_lagr} is
\begin{equation}
    \ten T\vec n = \vec t \qquad\text{on }\Gamma^{\mathrm{N}},
\end{equation}
where $\vec t = J^{-1} \| \ten F^T \vec n \| \vec t_0$ and the normal vectors are related by $\vec n = \| \ten F^{-T}\vec N \|^{-1} \ten F^{-T}\vec N$.

In the context of hyperelasticity, we postulate the existence of a strain energy density $\Psi=\Psi(\vec X,\,\ten F)$. Thus, by means of the Clausius-Duhem inequality, we obtain
\[
    \ten P = \ten P(\vec X,\,\ten F) = \frac{\partial \Psi}{\partial \ten F},\qquad P_{ij} = \frac{\partial\Psi}{\partial F_{ij}}.
\]
In the next section, we discuss the inverse counterpart of the problem described in this section, the so called \emph{inverse elasticity problem} (IEP).

\subsection{The inverse elasticity problem}
\label{sec:inverse_problem}
In solid mechanics, the DEP consists in reconstructing the current configuration given the reference configuration by solving \eqref{eq:balance_linear}, complemented by appropriate boundary conditions and by the constitutive assumptions on the material response.
In what follows we are interested in the IEP instead: the reconstruction of the relaxed configuration $\Omega_0$ given the current configuration $\Omega$ and the external loads. Consider the inverse deformation $\tvchi = \vec\chi^{-1}$, so that $\vec X = \tvchi (\vec x)$ and the inverse displacement is defined similarly to $\vec u$ as $\tu (\vec x)= \tvchi(\vec x) - \vec x$. The fields $\vec u$ and $\tu$ are related through the deformation fields as
\[
    \vec u = -\tu\circ \vec{\chi},\qquad \tu=-\vec{u}\circ\tvchi.
\]

\begin{remark}
    In what follows, for the sake of conciseness, we will omit the composition with $\vec{\chi}$ and $\tvchi$, when this will be clear from the context, and we will simply write, for instance, $\vec u = -\tu$.
\end{remark}

\begin{remark}
    \label{rem:stress_free}
    Unless specified differently, we restrict our attention to the situation in which the \emph{reference configuration} coincides with the \emph{relaxed configuration}, namely if $\ten F$ is the identity $\ten I$, we have
    \[
        \ten P(\vec X,\,\ten I) = \ten 0.
    \]
    We highlight that this might be restrictive, especially for living tissues, for which a relaxed configuration might not exist. Indeed, a variety of processes, such as growth~\cite{rodriguez1994stress,dicarlo2002growth,Goriely_2005}, active phenomena~\cite{kondaurov1987finite,taber2000modeling,Ambrosi_2011,Riccobelli_2019}, and plastic deformations~\cite{kroner1959allgemeine,Lee1969}, might produce local distortions that are geometrically incompatible. This leads to the generation of a stress state in the body even in the absence of external loads. The correct identification of the relaxed state of elastic bodies subject to these phenomena require specific treatments which go beyond the scope of the present article. Nonetheless, a remarkable case in which the theory described in this Section can be directly applied is the active stress approach, a method usually exploited to model contractility in muscle tissue. Such aspects are treated in Section~\ref{sec:active_stress}.
\end{remark}

We denote  the inverse deformation gradient by $\tF=\grad\tvchi=\ten I + \grad\tu$. The two deformation gradient tensors are related by $\tF = \ten F^{-1}$.

From \eqref{eq:balance_linear_cauchy}, the IEP can be cast as finding $\tu$ such that
\begin{equation}
    \label{eq:inverse}
    \left\{
    \begin{aligned}
         & \diver \ten{T}(\vec{x},\,\tF^{-1}) + \vec b=\vec 0 &  & \text{in }\Omega,              \\
         & \ten{T}(\vec{x},\,\tF^{-1})\vec n =\vec t          &  & \text{on }\Gamma^{\mathrm{N}}, \\
         & \tu = - \vec u_{\mathrm{D}}                        &  & \text{on }\Gamma^{\mathrm{D}}.
    \end{aligned}
    \right.
\end{equation}
It is well known that the direct problem has a variational structure, where the displacement field must minimize the functional
\begin{equation}
    \label{eq:functional_direct}
    \mathcal{F}[\vec u]=\int_{\Omega_0}\Psi(\vec{X},\,\ten F)\,\dX-\int_{\Gamma^{\mathrm{N}}_0}\vec{t}_0\cdot\vec u\,\d S-\int_{\Omega_0}\vec{B}\cdot\vec u\,\dX,
\end{equation}
where $\vec{t}_0$ and $\vec{B}$ are assumed to be function of $\vec X$ only. We shall also assume that $\Psi$ is polyconvex, namely there exists a convex function $g:\lin^+(\mathbb{R}^3)\times\lin^+(\mathbb{R}^3)\times \mathbb{R}^+\rightarrow \mathbb{R}\cup\{+\infty\}$ such that
\[
    \Psi(\ten F) = g(\ten F,\,\cof\ten F,\,J).
\]
This condition,  plus some growth conditions~\cite{ball1976convexity}, guarantee that the DEP \eqref{eq:balance_linear}-\eqref{eq:BC_lagr} admits a solution represented by a minimum of the functional \eqref{eq:functional_direct}.  In what follows, we will show that the situation is more complex for the inverse problem \eqref{eq:inverse}.

\subsubsection{Shield's transformation and convexity properties}\label{section:eshelby}

Under the assumption of material homogeneity, the variational structure of the problem follows from \emph{Shield's transformation}~\cite{Shield_1967}. Such a transformation leads to an equivalent formulation of \eqref{eq:balance_linear_cauchy}, where the Eshelby stress tensor $\eshelby$ takes the place of the Cauchy stress $\cauchy$. As noticed by Chadwick~\cite{Chadwick_1975}, such a correspondence suggests a duality between the Eshelby and the Cauchy stress tensors. In this section, we expand Shield's initial findings to encompass inhomogeneous materials, employing a methodology akin to that elucidated in Ref.~\cite{Merodio_2006}. A simple change of variable shows that
\[
    \int_{\Omega_0}\Psi(\vec X,\,\ten{F})\,\dX = \int_\Omega J^{-1} \Psi(\vec\chi^{-1}(\vec x),\,\ten{F})\,\dx.
\]
Thus, we introduce the dual energy density $\tPsi=\tPsi(\vec{x},\,\tF)$, defined as
\begin{equation}
    \label{eq:Shield}
    \tPsi(\vec{x},\,\tF)=\cur{J}\Psi(\tvchi(\vec x),\,\tF^{-1}).
\end{equation}
Equation \eqref{eq:Shield} is the Shield transformation of $\Psi$. We then introduce the spatial \emph{Eshelby stress} $\ten \Sigma$, defined as
\begin{equation}
    \label{eq:eshelby}
    \ten\Sigma \coloneqq \frac{\partial \tPsi}{\partial \tF}=\cur{J}\tF^{-T}\left(\Psi\ten I-\ten P\tF^{-T}\right) = \cur{\ten{F}}^{-T}\left( \cur{\Psi}\ten{I}-\ten{T}\right).
\end{equation}
Since
\[
    \frac{\d }{\d x_i} \tPsi (\vec{x},\,\tF(\vec{x})) = \frac{\partial\tPsi}{\partial x_i} +\sum_{h,k=1}^3\frac{\partial\tPsi}{\partial \cur{F}_{hk}}\frac{\partial \cur{F}_{hk}}{\partial x_i}=\frac{\partial\tPsi}{\partial x_i} +\sum_{h,k=1}^3\Sigma_{hk}\frac{\partial \cur{F}_{hk}}{\partial x_i},
\]
and
\[
    \begin{aligned}
        \frac{\partial}{\partial x_j}\left(\sum_{h=1}^3\cur{F}_{hi}\Sigma_{hj}\right) & = \sum_{h=1}^3\left(\frac{\partial \cur{F}_{hi}}{\partial x_j}\Sigma_{hj} + \cur{F}_{hi}\frac{\partial\Sigma_{hj}}{\partial x_j}\right)=                                   \\
                                                                                      & =\sum_{h=1}^3\left(\frac{\partial \tPsi}{\partial \cur{F}_{hj}}\frac{\partial \cur{F}_{hi}}{\partial x_j} + \cur{F}_{hi}\frac{\partial\Sigma_{hj}}{\partial x_j}\right)=   \\
                                                                                      & =\sum_{h=1}^3\left(\frac{\partial \tPsi}{\partial \cur{F}_{hj}}\frac{\partial \cur{F}_{hj}}{\partial x_i} + \cur{F}_{hi}\frac{\partial\Sigma_{hj}}{\partial x_{j}}\right),
    \end{aligned}
\]
from \eqref{eq:eshelby} we can rewrite the momentum equation in terms of the Eshelby stress tensor:
\begin{equation}
    \label{eq:divTdivSigma}
    \begin{aligned}
        \diver \cauchy & = \diver(\tPsi\ten{I}) -\diver(\tF^T\eshelby) =                                                  \\
                       & =\grad\tPsi -\diver(\tF^T\eshelby) =                                                             \\
                       & =\frac{\partial\tPsi}{\partial\vec x}+\eshelby:\grad\tF - \eshelby:\grad\tF-\tF^T\diver\eshelby= \\
                       & =\frac{\partial\tPsi}{\partial\vec x}-\tF^T\diver\eshelby,
    \end{aligned}
\end{equation}
where $\partial\tPsi/\partial\vec x$ is the partial derivative of $\tPsi(\vec x,\,\tF(\vec x))$ with respect to its first argument. Thus, the problem \eqref{eq:inverse} is equivalent to the following one expressed in terms of the Eshelby stress tensor
\begin{equation}
    \label{eq:balance_eshelby}
    \left\{
    \begin{aligned}
         & \diver \eshelby + \vec \beta = \vec 0 &  & \text{in }\Omega,              \\
         & \eshelby\vec{n}=\vec\sigma            &  & \text{on }\Gamma^{\mathrm{N}}, \\
         & \tu = - \vec u_{\mathrm{D}}           &  & \text{on }\Gamma^{\mathrm{D}},
    \end{aligned}
    \right.
\end{equation}
where, from \eqref{eq:eshelby} and \eqref{eq:divTdivSigma}, we have
\begin{equation}
    \label{eq:betasigma}
    \vec\beta = -\tF^{-T}\left(\vec b+\frac{\partial \tPsi}{\partial\vec x}\right),\qquad\vec\sigma = \tF^{-T}(\tPsi\vec n - \vec t).
\end{equation}

We remark that such a formulation holds for any constitutive assumptions and, up to our knowledge, it has not been reported elsewhere. Previous derivations all assume that the body is homogeneous, i.e. $\tPsi$ does not depend on $\vec x$. This is relevant for anisotropic materials, and more so in cardiac mechanics since the direction of the fibers usually depends on $\vec x$, rendering the term $\vec{\beta}$ in \eqref{eq:balance_eshelby} not zero even in the absence of body forces. %
We have summarized the main quantities involved in the direct and the inverse formulations in Table~\ref{table:forms-summary}.

In the particular case of homogeneous materials ($\partial \tPsi/\partial \vec{x}=0$), problem \eqref{eq:balance_eshelby} can be written as a minimization problem assuming that $\vec \beta$ and $\vec \sigma$ are measurable functions depending only on space:
\begin{equation}
    \label{eq:hyp_beta_sigma}
    \vec\beta: \Omega\rightarrow\mathbb{R}^3,\quad\vec\sigma: \Omega\rightarrow\mathbb{R}^3.
\end{equation} Then, under these assumptions, it can be seen that \eqref{eq:balance_eshelby} is equivalent to finding the stationary points of the functional
\begin{equation}
    \label{eq:functional_eshelby}
    \widehat{\mathcal{F}}[\widehat{\vec u}]=\int_{\Omega}\tPsi(\tF)\,\dx-\int_{\Gamma^{\mathrm{N}}}\vec{\sigma}(\vec{x})\cdot\widehat{\vec u}(\vec{x})\,\d S-\int_{\Omega}\vec{\beta}(\vec{x})\cdot\widehat{\vec u}(\vec x)\,\dx.
\end{equation}
Remarkably, Shield's transformation \eqref{eq:Shield} preserves polyconvexity or rank-1 convexity if $\Psi$ is polyconvex or rank-1 convex as well, see Proposition 17.6.2 in Reference~\cite{_ilhav__1997}. Ball's theorem on the existence of energy minimizers~\cite{ball1976convexity} can then be used to prove the existence of minimizers \eqref{eq:functional_eshelby}.

However, in practical applications, this is almost never the case. Indeed, the existence theorem can be applied if $\vec\beta$ and $\vec\sigma$ are functions of $\vec x$ only, as assumed in \eqref{eq:hyp_beta_sigma}, but, in most applications, $\vec\beta$ and $\vec\sigma$  depend on $\tF$ as well, and \eqref{eq:balance_eshelby} is not anymore equivalent to finding the stationary points of \eqref{eq:functional_eshelby}. This aspect can create some issues as shown in the following two examples.

\begin{table}
    \centering
    \begin{tabular}{@{}c  c  c  c@{}}
        \toprule
        Quantity     & Direct           & Inverse (Cauchy)                                   & Inverse (Eshelby)                                                           \\ \midrule
        Energy       & $\Psi(\ten F)$   & --                                                 & $\tPsi(\tF)= \cur{J}\Psi(\tF^{-1})$                                         \\
        Stress       & $\ten P(\ten F)$ & $\ten T(\tF^{-1})=\cur{J}\ten P(\tF^{-1})\tF^{-T}$ & $\ten \Sigma(\tF)=\tF^{-T}\left(\tPsi(\tF)\ten I - \ten T(\tF^{-1})\right)$ \\
        Volume load  & $\vec B$         & $\vec b=\cur{J}\vec B$                             & $\vec \beta =  -\tF^{-T}\left(\vec b+\partial \tPsi/\partial\vec x\right)$  \\
        Surface load & $\vec t_0$       & $\vec t=\cur{J}\|\ten F^T\vec n\|\vec t_0$         & $\vec \sigma =\tF^{-T}(\tPsi\vec n - \vec t)$                               \\\bottomrule
    \end{tabular}
    \caption{Energy, stress, and load terms in all three formulations: DEP \eqref{eq:balance_linear}, IEP in terms of the Cauchy stress \eqref{eq:inverse} and of the Eshelby stress \eqref{eq:balance_eshelby}. We omit the explicit dependence on $\vec X$ and $\vec x$.}
    \label{table:forms-summary}
\end{table}
\subsubsection{Elastic disk subject to an external pressure: non-existence of radially symmetric solutions}
\begin{figure}[ht!]
    \centering
    \includegraphics[width=0.5\textwidth]{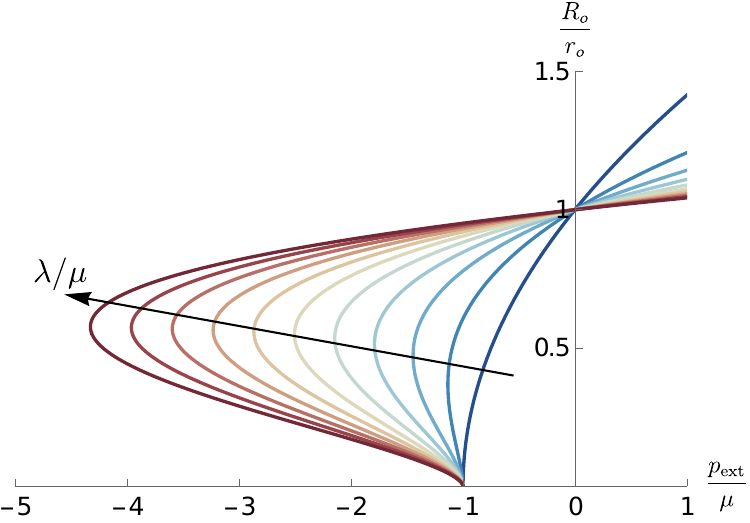}
    \caption{Plot of $R_o/r_o$ as a function of $p_\text{ext}/\mu$ for $\lambda/\mu=0,\,1,\dots,\,10$. The arrow denotes the direction in which $\lambda/\mu$ increases.}
    \label{fig:Ro}
\end{figure}

Consider now the circular domain $\Omega=B(O,\,r_o)\subset\mathbb{E}^2$ representing the current configuration, where $B(O,\,r_o)$ is the disk of center $O$ and radius $r_o$. Let $(R,\,\Theta)$ and $(r,\,\theta)$ be the reference and the spatial polar coordinates of a generic point about the origin $O$. We assume that the sphere is subject to a pressure $p_\text{ext}$, so that
\begin{equation}
    \label{eq:BC_p}
    \ten T\vec e_r = - p_\text{ext}\vec e_r,
\end{equation}
where $(\vec{e}_R,\,\vec{e}_\Theta)$ and $(\vec{e}_r,\,\vec{e}_\theta)$ are the polar basis in the reference and in the current configuration, respectively.
We assume that the material behaves as a compressible Neo-Hookean material, given by a strain energy density defined as
\begin{equation}\label{eq:neo-hookean}
    \Psi (\ten F) = \frac{\mu}{2}(\tr (\ten F^T\ten F) - 2 \log J - 2)+\frac{\lambda}{2}(\log J)^2,
\end{equation}
where $\lambda$ and $\mu$ are the (linear) Lame's parameters.
Under such assumptions, the Cauchy stress tensor reads
\begin{equation}
    \label{eq:Cauchy_NH}
    \ten T = \frac{1}{J}\left(\mu\ten F \ten F^T + (\lambda \log J - \mu)\ten I\right).
\end{equation}
We assume polar symmetry, so that $r=r(R)$ and $\theta=\Theta$. The deformation gradient is given by
\[
    \ten{F}=r'\vec{e}_r\otimes\vec{e}_R+\frac{r}{R}\vec{e}_\theta\otimes\vec{e}_\Theta.
\]
Due to the symmetries of the deformation field, the balance of linear momentum \eqref{eq:balance_linear_cauchy} reduces to the following ordinary differential equation
\begin{equation}
    \label{eq:lin_mom_polar}
    \frac{\d T_{rr}}{\d r} + \frac{T_{rr}-T_{\theta\theta}}{r}=0.
\end{equation}
We observe that $r=r_o R/R_o$ satisfies \eqref{eq:lin_mom_polar}. Here, $R_o\in\mathbb{R}$ is the reference radius of the disk that can be found by enforcing the boundary condition \eqref{eq:BC_p}, obtaining
\begin{equation}
    \label{eq:Ro_disk}
    \frac{R_o ^2}{r_o^2} \lambda  \log \left(\frac{r_o^2}{R_o ^2}\right)+\left(1 - \frac{R_o ^2}{r_o^2}\right) \mu  = - p_\text{ext},
\end{equation}
whose solution can be expressed as
\[
    R_o^2=\left(\mu +p_\text{ext}\right)\left(\lambda  W_0\left(\frac{e^{\mu /\lambda } (\mu +p_\text{ext})}{\lambda }\right)\right)^{-1}r_o^2,
\]
where $W_0$ is the principal branch of the Lambert function ($w=W_0(z)$ is the solution of $w e^w=z$, with $z$ being a complex number), see Fig.~\ref{fig:Ro}. In the special case $\lambda=0$, the solution of \eqref{eq:Ro_disk} is given by
\[
    R_o^2 = \frac{\mu+p_\text{ext}}{\mu}.
\]

We observe that depending on the value of the applied pressure, the inverse problem may not have solutions with radial symmetry.
In particular, we observe that in \eqref{eq:Ro_disk} as $R_o\rightarrow0$ we have $
    p_\text{ext}\rightarrow-\mu$
and for all the values of $\lambda$. This is a limit case where the reference configuration shrinks to a single point for a finite value of the external pressure. Thus, the IEP might have no solution if we apply constant Neumann boundary conditions and if we choose a polyconvex strain energy density.

\subsection{Injectivity of the inverse deformation}
\label{sec:self-penetration}
\begin{figure}[b!]
    \centering
    \includegraphics[width=0.8\textwidth]{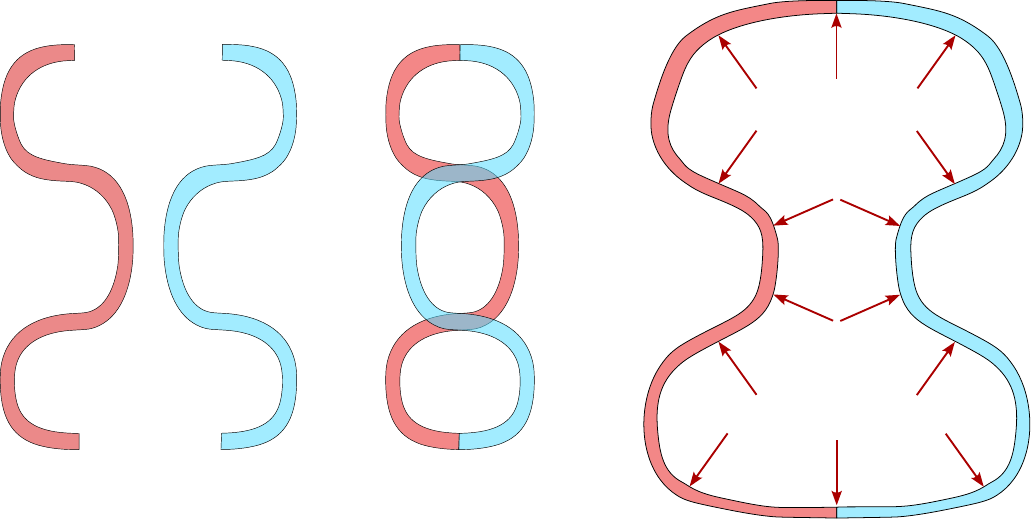}
    \caption{Left: representation of two symmetric pieces in their relaxed configuration. Center: relaxed state of the two pieces after they are glued together; the relaxed state of the body exhibits a self-intersection; Right: deformed configuration of the body subject to an inner pressure.}
    \label{fig:self_intersection}
\end{figure}

Usually, the deformation field is supposed to be bijective to avoid self-intersections of the body. However, requiring that $\tvchi$ be bijective might be too strict for the free body problem. Indeed, if we take two S-shaped pieces, we can imagine to glue them together, as shown in Fig.~\ref{fig:self_intersection}, and the relaxed state of the body requires a self-intersection. This situation is not uncommon and, as shown in the following paragraphs, applies to the heart as well.

Thus, the relaxed configuration of the body cannot in general be achievable in reality due to a global geometric incompatibility, i.e. $\tvchi$ is not injective. The analysis proposed in this section is still valid since $\vec{\chi}$ is locally invertible due to the inverse function theorem ($J=\det\Grad\vec{\chi}>0$).

Some issues may arise if we want to describe a further deformation of $\Omega$ to $\overline{\Omega}$ with $\overline{\Omega}=\overline{\vec \chi}(\Omega)$. In such a case, the relaxed configuration $\Omega_0$ cannot be used as a reference configuration due to the self-intersection of the body. A possible strategy to solve such a problem is to take $\Omega$ as a reference configuration. In such a case, similarly to the multiplicative decomposition of the deformation gradient exploited to model plasticity~\cite{kroner1959allgemeine, Lee1969}, growth and remodelling~\cite{rodriguez1994stress,epstein2015mathematical}, we introduce
\[
    \ten{F}_\text{e} = \overline{\ten{F}}\tF^{-1},
\]
where $\overline{\ten F}=\grad\overline{\vec \chi}$ and $\ten{F}_\text{e}$ is the elastic distortion from the relaxed state to the configuration $\overline{\Omega}$. Then, the strain energy density per unit volume of $\Omega$ is given by
\[
    \psi(\overline{\ten{F}}) = (\det{\tF})\Psi(\ten{F}_\text{e}).
\]
This underlines the possibility to treat local and global geometric incompatibilities in a unique way. The local incompatibilities manifest themselves as a non-compatible $\tF$, i.e. there does not exist a function $\tvchi$ such that $\tF = \grad \tvchi$~\cite{riccobelli2019existence}, while a global geometric incompatibility is a non-injectivity of $\tvchi$. The inclusion of local geometric incompatibilities of the relaxed state in the IEP is left as a possible future study.

\subsection{Active stress}
\label{sec:active_stress}
When modeling biological tissues, and the myocardium in particular, it is important to take into account the active forces that are involved during muscle contraction. One of these approaches is the so called \emph{active stress}. Usually it is assumed that there exists a reference configuration $\Omega_0$ that is \emph{stress free in the abscence of active forces}~\cite{Ambrosi_2011,Riccobelli_2019}. Let $\Psi_\text{pas}(\vec{X},\,\ten{F})$, be the strain energy density of the passive material, and we introduce the passive first Piola-Kirchhoff stress tensors, defined as
\begin{equation}
    \label{eq:Ppas_e_psipas}
    \ten{P}_\text{pas}(\vec X,\,\ten F) \coloneqq \frac{\partial \Psi_\text{pas}}{\partial \ten F}.
\end{equation}
We require $\ten{P}_\text{pas}$ to satisfy
\[
    \ten{P}_\text{pas}(\vec X,\,\ten I)=\ten 0\qquad\forall\vec X\in\Omega_0.
\]
Let $\ten{P}_\text{act}(\vec X,\,\ten{F};\,T_a)$ be a tensor-valued function representing the active stress generated by the muslce fiber contractility. Here, $T_a$ is a parameter describing the tension generated by the muscle, which is zero in the passive case.
Hence, we assume that $\ten{P}_\text{act}(\vec X,\,\ten F;\,0)=\ten{0}$. The active stress approach envisages writing first Piola--Kirchhoff stress tensor as
\begin{equation}
    \label{eq:act_stress}
    \ten{P} = \ten{P}_\text{pas}+\ten{P}_\text{act}.
\end{equation}
In conclusion, when considering active materials, the IEP shall be regarded as that of finding the configuration assumed by the body in the absence of both passive and active stress, that is when $\ten{P}_\text{pas} = \ten{P}_\text{act} = \ten 0$.

\subsubsection{Active stress in cardiac mechanics problems}

As it is standard in the cardiac modeling literature, we consider an orthonormal triplet $(\fiberRefF,\,\fiberRefS,\,\fiberRefN)$ of fibers, sheets, sheet-normal directions~\cite{piersanti2021modeling,bayer2012novel}.
The fiber architecture plays a role in determining both the passive and the active response of the tissue. A common choice for the active stress tensor is
\begin{equation}
    \label{eq:Pact}
    \ten{P}_\text{act} = S_f(\|\ten{F}\fiberRefF\|; T_a)\frac{\ten F\fiberRefF\otimes\fiberRefF}{\|\ten{F}\fiberRefF\|} + S_n(\|\ten{F}\fiberRefN\|;\,T_a)\frac{\ten F\fiberRefN\otimes\fiberRefN}{\|\ten{F}\fiberRefN\|},
\end{equation}
where $S_f$ and $S_n$ are scalar functions. We remark that, if $S_f$ and $S_n$ are integrable with respect to $\|\ten{F}\fiberRefF\|$ and $\|\ten{F}\fiberRefN\|$, respectively, and $\psi_f$ and $\psi_n$ are their primitives, we have~\cite{Giantesio_2018,regazzoni2021oscillation}$^{,\,}$\footnote{As shown for skeletal muscles where a single family of fiber is present~\cite{Giantesio_2018}, such an approach is equivalent to model the tissue as a mixture of passive and active elements, for details see Ref.~\cite{Riccobelli_2019}.}
\begin{equation}
    \label{eq:psifs}
    \ten{P}_\text{act} = \frac{\partial \psi_f}{\partial \ten{F}} + \frac{\partial \psi_s}{\partial \ten{F}} = \frac{\psi_f'(\|\ten{F}\fiberRefF\|; T_a)}{\|\ten{F}\fiberRefF\|}\ten F\fiberRefF\otimes\fiberRefF + \frac{\psi_n'(\|\ten{F}\fiberRefN\|;\,T_a)}{\|\ten{F}\fiberRefN\|}\ten F\fiberRefN\otimes\fiberRefN,
\end{equation}
where $'$ denotes the differentiation with respect to the first argument.
Thus, the total (passive and active) stress tensor can be associated with the strain energy density:
\begin{equation}
    \label{eq:psi_active_stress}
    \Psi(\vec X,\,\ten F;\,T_a) = \Psi_\text{pas}(\ten F) + \Psi_f(\|\ten{F}\fiberRefF\|; T_a) + \Psi_n(\|\ten{F}\fiberRefN\|;\,T_a).
\end{equation}
In light of this observation, the procedure exposed in Section~\ref{section:eshelby} can be applied to the energy $\psi$ defined in \eqref{eq:psi_active_stress}.

\section{Numerical approximation}
\label{section:numerical-approximation}
In this section, we provide the details to solve numerically equation \eqref{eq:inverse}. For this, we provide (i) the weak formulation, (ii) a detailed description of the IEP formulation for cardiac modeling, and (iii) a simple implementation of the IEP to show that an existing DEP solver can be turned into an IEP solver with little modifications.

\subsection{Weak formulation}
The weak formulation of the DEP \eqref{eq:balance_linear} can then be stated as finding $\vec u \in V_0^{\vec u_{\mathrm{D}}}$ such that
\begin{equation}\label{eq:forward weak}
    \int_{\Omega_0} \ten P(\vec X,\,\ten F):\Grad \vec v\, \dX + \int_{\Gamma_0^N}\vec{t}_0\cdot \vec v\,\d S = \int_{\Omega_0}\vec{B}\cdot\vec v\,\dX, \qquad \forall \vec v\in V_0,
\end{equation}
where we have defined the trial and test function spaces:
\begin{equation*}
    \begin{split}
        V_0^{\vec u_{\mathrm{D}}}   &= \{ \vec v \in H^1(\Omega_0; \mathbb{R}^3) \text{ s.t. } \vec v = \vec u_{\mathrm{D}} \text{ on } \Gamma^{\mathrm{D}}_0\}, \\
        V_0              &= \{ \vec v \in H^1(\Omega_0; \mathbb{R}^3) \text{ s.t. } \vec v = \vec 0   \text{ on } \Gamma^{\mathrm{D}}_0\}, \\
    \end{split}
\end{equation*}
and the weak formulation of the IEP can be stated as finding $\tu \in V^{\vec u_{\mathrm{D}}}$ such that
\begin{equation}\label{eq:backward weak}
    \int_{\Omega} \ten T(\vec x,\,\tF^{-1}):\grad \vec v\,\dx + \int_{\Gamma^N}\vec t\cdot \vec v\,\d s = \int_{\Omega}\vec b\cdot\vec v\,\dx, \qquad \forall \vec v\in V,
\end{equation}
where
\begin{equation*}
    \begin{split}
        V^{\vec{u}_{\mathrm{D}}}   &= \{ \vec v \in H^1(\Omega; \mathbb{R}^3) \text{ s.t. } \vec v = -\vec u_{\mathrm{D}} \text{ on } \Gamma^{\mathrm{D}}\}, \\
        V              &= \{ \vec v \in H^1(\Omega; \mathbb{R}^3) \text{ s.t. } \vec v = \vec 0   \text{ on } \Gamma^{\mathrm{D}}\}, \\
    \end{split}
\end{equation*}
and we recall that
\[
    \ten T(\vec x,\,\tF^{-1}) = \cur{J}\ten P(\tvchi(\vec x),\,\tF^{-1})\tF^{-T}.
\]
Similarly, an equivalent weak formulation of \eqref{eq:backward weak} using the strong formulation \eqref{eq:balance_eshelby} reads
\begin{equation}\label{eq:eshelby_weak}
    \int_{\Omega} \ten \Sigma(\vec x,\,\tF):\grad \vec v\dx + \int_{\Gamma^N}\vec \sigma\cdot \vec v\d s = \int_{\Omega}\vec \beta\cdot\vec v\dx, \qquad \forall \vec v\in V,
\end{equation}
where the Eshelby stress tensor $\ten \Sigma$ is defined in \eqref{eq:eshelby}.

\subsection{Cardiac inverse model} \label{sec:cardio_model}

In this section, we derive the IEP in the setting of cardiac modeling.
Specifically, we account for the presence of an active stress and of cardiac fibers, and we consider boundary conditions often used to account for the interactions of the heart with the blood and with the surrounding organs.

\subsubsection{The direct problem}

\paragraph{{Geometry}}
Let $\Omega_0$ be the stress-free configuration of the {passive} myocardium, and $\Omega$ the deformed configuration.
We include in the domain also segments of the main vessels connecting the heart to the circulatory system (aorta, pulmonary artery and main veins).
Typically, geometries available from medical imaging are acquired at diastasis, namely one of the last phases of diastole, right before the atrial kick (the beginning of atrial systole).
This phase of the heartbeat is the one in which the heart is most stationary, thus facilitating the medical imaging acquisition process.
    {Furthermore, being inertial forces negligible, a quasi-static assumption is well motivated at this stage.}
At diastasis, the blood pressures in the four chambers are relatively small, compared to the rest of the heartbeat, and active forces are also small.
These features facilitate the IEP resolution.

\paragraph{{Constitutive assumptions}}

We use the active stress approach described in Section~\ref{sec:active_stress} to model myocardium contractility. Specifically, we adopt the active stress tensor as in \eqref{eq:Pact}-\eqref{eq:psifs}, where we choose~\cite{regazzoni2022emcirculation}
\begin{equation}
    \label{eq:const_psisf}
    \psi_f(\|\ten F \fiberRefF\|) = T_a \|\ten F \fiberRefF\|,\quad\psi_s(\|\ten F \fiberRefS\|) = \ratioN T_a \|\ten F \fiberRefS\|,
\end{equation}
where $\TensAct$ denotes the active tension, acting mainly in the direction of fibers $\fiberRefF${. The fibers are not perfectly aligned due to fiber dispersion. This is modeled through the introduction of the constant parameter $0<\ratioN<1$ in \eqref{eq:const_psisf}.}
The Piola-Kirchhoff stress tensor thus reads~\cite{Giantesio_2018,regazzoni2022emcirculation}
\begin{equation*}
    {\ten P
        = \frac{\partial \Psi_\text{pas}(\vec X,\,\ten{F})}{\partial \ten F}
        + \TensAct
        \left[
            \frac{\ten F \fiberRefF \otimes \fiberRefF}{\| \ten F \fiberRefF \|}
            + \ratioN
            \frac{\ten F \fiberRefN \otimes \fiberRefN}{\| \ten F \fiberRefN \|}
            \right],}
\end{equation*}
We remark that, while solving the IEP for cardiac models, the active stress term is often neglected.
However, it is important to notice that in any moment of the heartbeat (even at diastasis), a non-negligible amount of active tension is present, known as diastolic tension~\cite{katz2010,regazzoni2019reviewXB}.
Hence, it is crucial to account for the active stress during the stress-free configuration recovery procedure.

    {As a constitutive choice for the passive contribution to the strain energy density (see \eqref{eq:Ppas_e_psipas}-\eqref{eq:act_stress}), we use a function $\Psi_\text{pas}(\vec{X},\,\ten F)$, where the explicit dependence on $\vec{X}$ is necessary to account for the anisotropic behaviour induced by the presence of muscle fibers. We use different expressions for $\Psi_\text{pas}$, which are explicitly specified in what follows.}

\paragraph{{Boundary conditions}}
We split the boundary $\partial\Omega_0$ of the domain in different subsets, and apply boundary conditions depending on the interacting tissues within each subset.
The internal boundaries of the myocardium are in contact with blood, which exerts a pressure on the myocardium.
We consider $\numCavities = 6$ cavities (namely the four cardiac chambers, the aorta and the pulmonary artery), in which the blood pressure can be reasonably considered constant.
    {For} each cavity $i$ (with $i = 1,\dots,\numCavities$), {we denote its boundary in the reference configuration by $\bdryEndoREF{i} \subset \partial \Omega_0$.}
    {We model the action of the blood on the cavity surfaces as a constant hydrostatic pressure $p_i$}:
\begin{equation*}
    \ten P\vec{N} = -p_i J \ten F^{-T} \vec{N}
    \quad \text{on $\bdryEndoREF{i}$}.
\end{equation*}
The epicardium, that is the external surface of the heart, is instead in contact with the pericardium, a tough fibroelastic sac containing the heart and the roots of the great vessels.
    {We} model the interaction of the heart with the pericardium by applying (anisotropic) linear springs on the pericardial surface $\bdryEpiREF$~\cite{regazzoni2022emcirculation,pfaller2019importance}:
\begin{equation}\label{eqn:BC_epicardium}
    \ten P\vec{N}
    {= -} \BCKn (\vec{N} \otimes \vec{N}) \vec u
        {-} \BCKt (\ten I - \vec{N} \otimes \vec{N}) \vec u
    \quad \text{on $\bdryEpiREF$},
\end{equation}
where the positive coefficients $\BCKn$ and $\BCKt$ account for the elastic response of the pericardium and the surrounding organs in the normal and tangent direction, respectively.

Finally, we apply homogeneous {Dirichlet} boundary conditions on the artificial boundaries originating where arteries and veins are truncated, which we denote by $\bdryDREF$.

\paragraph{{Weak formulation}}
In conclusion, the weak formulation of the DEP consists in finding
$\vec u \in
    V_0 = \{
    \vec v \in H^1(\Omega_0; \mathbb{R}^3)
    \text{ s.t. }
    \vec v = \vec 0 \text{ on $\bdryDREF$}
    \}$
such that
\begin{equation}\label{eq:direct weak cardio}
    \begin{split}
        \int_{\Omega_0} \ten P(\ten F):\Grad \vec v\dx
        =&-\sum_{i = 1}^{\numCavities} \int_{\bdryEndoREF{i}} p_i J \ten F^{-T} \vec{N} \cdot \vec v\,\d s
        \\
        &- \int_{\bdryEpiREF} \BCKt \vec u \cdot \vec v \,\d s
        \\
        &- \int_{\bdryEpiREF} (\BCKn - \BCKt)
        \left( \vec N \cdot \vec u   \right)
        \left( \vec N \cdot \vec v \right)\,\d s
    \end{split}
\end{equation}
for all $\vec{v}\in V_0$.
\subsubsection{The inverse problem}

By proceeding as above, we derive the following IEP formulation for the cardiac model: we look for
$\tu \in
    V = \{
    \vec v \in H^1(\Omega; \mathbb{R}^3)
    \text{ s.t. }
    \vec v = \vec 0 \text{ on $\bdryDACT$}
    \}$
such that
\begin{equation}\label{eq:backward weak cardio}
    \begin{split}
        \int_{\Omega} \cur{J} \ten P(\tF^{-1})\tF^{-T}&:\grad \vec v\dx
        =-\sum_{i = 1}^{\numCavities} \int_{\bdryEndoACT{i}} p_i \vec n\cdot \vec v\,\d s
        \\
        &+ \int_{\bdryEpiACT} \BCKt \cur{J} \|\tF^{-T} \vec n\| \tu \cdot \vec v \,\d s
        \\
        &+ \int_{\bdryEpiACT} (\BCKn - \BCKt) \frac{\cur{J}}{\|\tF^{-T} \vec n\|}
        \left( \tF^{-T} \vec n \cdot \tu   \right)
        \left( \tF^{-T} \vec n \cdot \vec v \right)\,\d s
    \end{split}
\end{equation}
for all $\vec v\in V$.

\subsection{Remarks on implementation}
In this section, we show that it is very simple to modify a solver for problem \eqref{eq:forward weak} to obtain a solver for problem \eqref{eq:backward weak}, at least when relying on an automatic differentiation engine. To show this, we will provide an example using the Unified Form Language (UFL)~\cite{Aln_s_2014}, but the concepts are still valid for other equivalent systems. We start by looking at how a simple formulation of nonlinear elasticity could look like in Listing~\ref{fig:ufl forward}, which can be found among the demos at the documentation of FEniCS~\cite{alnaes2015fenics}. We point out a similar implementation can be found in \cite{genet_2023_10299077}.

\begin{lstlisting}[language=python, frame=single, caption=UFL formulation of \eqref{eq:forward weak}., label={fig:ufl forward}, float]
V = VectorFunctionSpace(mesh, 'CG', 1)
u = Function(V)
v = TestFunction(V)
F = variable(grad(u) + Identity(3))  # Compute original one to diff
J = det(F)
Cbar = J**(-2/3) * F.T * F
E, nu = 1.0e4, 0.3
mu = Constant(E/(2*(1 + nu)))
lmbda = Constant(E*nu/((1 + nu)*(1 - 2*nu)))
psi = (mu / 2) * (tr(Cbar) - 3) + 0.5 * lmbda * (J-1) * ln(J)
P = diff(psi, F)
residual = inner(P, grad(v)) * dx - dot(Constant((0,0,-1)), v)* dx
bcs = DirichletBC(V, Constant((0,0,0)), "on_boundary") 
solve(residual==0, u, bcs=bcs)
    \end{lstlisting}
To convert this formulation, we need to (i) push forward the objects in the integrals and (ii) recast the kinematic quantities in terms of the inverse displacement. For this we have to observe that the Piola-Kirchhoff tensor is \emph{still} the derivative of $\Psi$ with respect to $\ten F$. This yields the formulation shown in Listing~\ref{fig:ufl backward}, where it can be seen that the difference between both codes is limited.

\begin{lstlisting}[language=python, frame=single, caption=UFL formulation of \eqref{eq:backward weak}., label={fig:ufl backward}, float]
V = VectorFunctionSpace(mesh, 'CG', 1)
u_hat = Function(V)
v = TestFunction(V)
F_hat = Identity(3) + grad(u_hat)  # Inverse tensor for inverse problem
J_hat = det(F_hat)
F = variable(inv(F_hat))  # Compute original one to differentiate
J = det(F)
Cbar = J**(-2/3) * F.T * F
E, nu = 1.0e4, 0.3
mu = Constant(E/(2*(1 + nu)))
lmbda = Constant(E*nu/((1 + nu)*(1 - 2*nu)))
psi = (mu / 2) * (tr(Cbar) - 3) + 0.5 * lmbda * (J-1) * ln(J)
P = diff(psi, F)
residual = (J_hat * inner(P, grad(v) * inv(F_hat)) * dx 
            - J_hat * dot(Constant((0,0,-1)), v)* dx)
bcs = DirichletBC(V, Constant((0,0,0)), "on_boundary") 
solve(residual==0, u_hat, bcs=bcs)
    \end{lstlisting}

\section{Self-intersection of the stress-free state}\label{section:physical-ref-conf}

\begin{figure}[ht!]
    \centering
    \begin{subfigure}[b]{0.45\textwidth}
        \includegraphics[width=\textwidth]{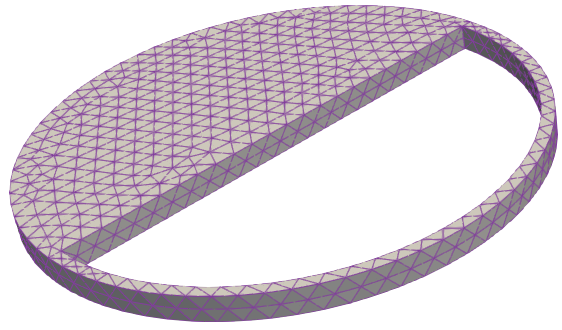}
        \caption{}
    \end{subfigure}
    \begin{subfigure}[b]{0.45\textwidth}
        \includegraphics[width=\textwidth]{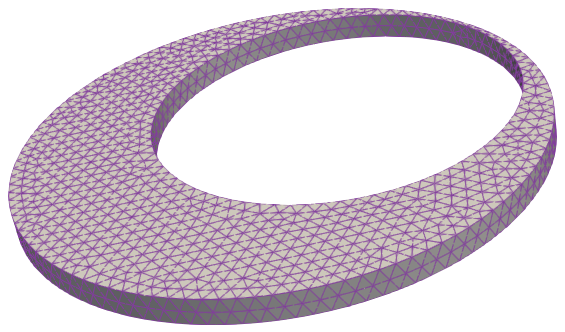}
        \caption{}
    \end{subfigure}
    \caption{(a) Semi-circle and (b) eclipse meshes used to study self-intersection mechanisms in the inverse displacement problem.}
    \label{fig:self-intersect-meshes}
\end{figure}

In this section, we discuss several aspects regarding the existence of a stress-free configuration. For this, we consider two simple geometries that represent a transverse cut of an idealized left ventricle as displayed in Figure~\ref{fig:self-intersect-meshes}. We refer to them as (a) the semi-circle and (b) the eclipse.

As discussed in Section~\ref{sec:self-penetration}, a global geometric incompatibility can result into self-intersecting relaxed states. We show two mechanisms under which this phenomenon can be seen, namely inner self-intersections and outer self-intersections. We show this in the presented geometries by considering the inner surface as an endocardium where a given pressure is known, and on the epicardium we consider the elastic response that arises from the interaction with the pericardium.

\begin{figure}[ht!]
    \begin{subfigure}{0.45\textwidth}
        \includegraphics[width=\textwidth]{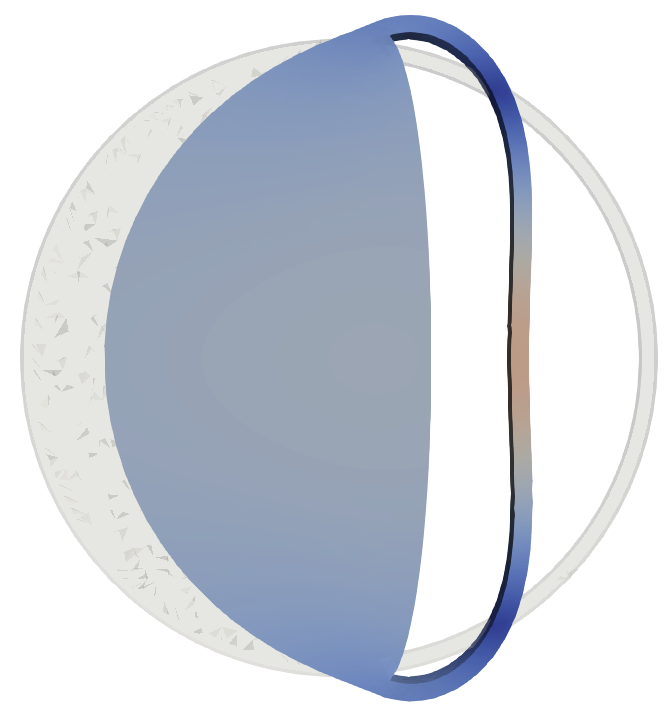}
        \caption{\SI{200}{\pascal}}
    \end{subfigure}
    \begin{subfigure}{0.45\textwidth}
        \includegraphics[width=\textwidth]{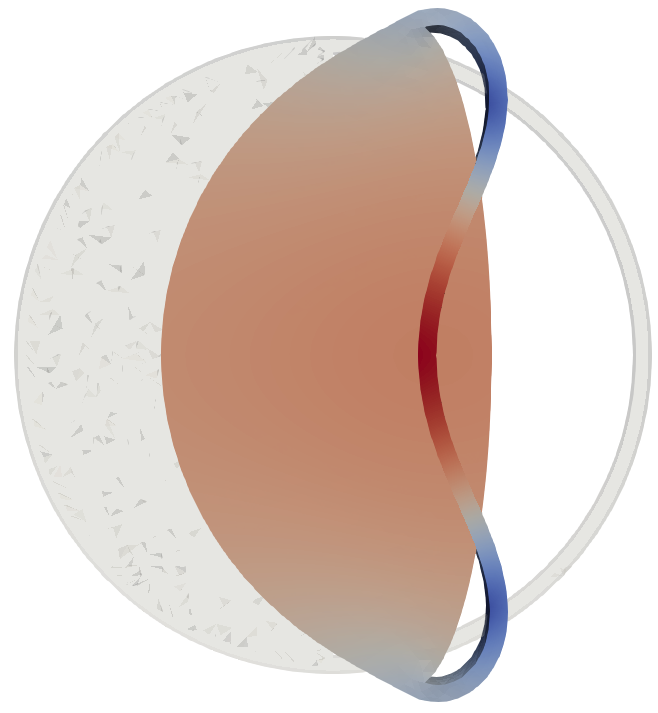}
        \caption{\SI{400}{\pascal}}
    \end{subfigure}
    \caption{Solution of the inverse displacement problem on the semi-circle geometry. The stress-free configuration is computed for (a) \SI{200}{\pascal} and (b) \SI{400}{\pascal}.}
    \label{fig:semicircle-solution}
\end{figure}

\begin{figure}[ht!]
    \begin{subfigure}{0.45\textwidth}
        \includegraphics[width=\textwidth]{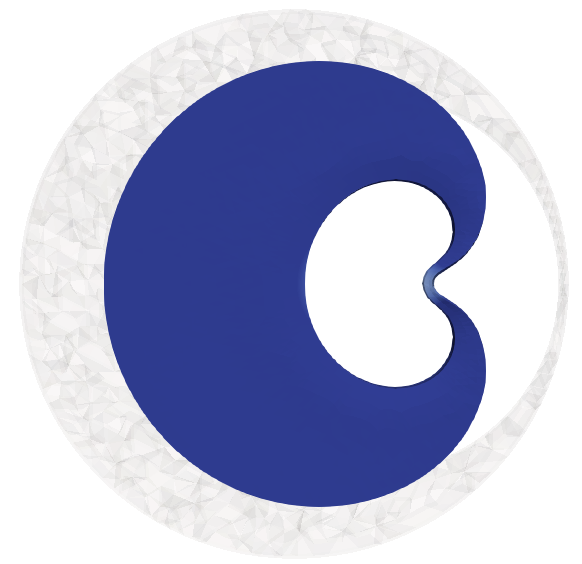}
        \caption{\SI{0.25e5}{\pascal}}
    \end{subfigure}
    \begin{subfigure}{0.45\textwidth}
        \includegraphics[width=\textwidth]{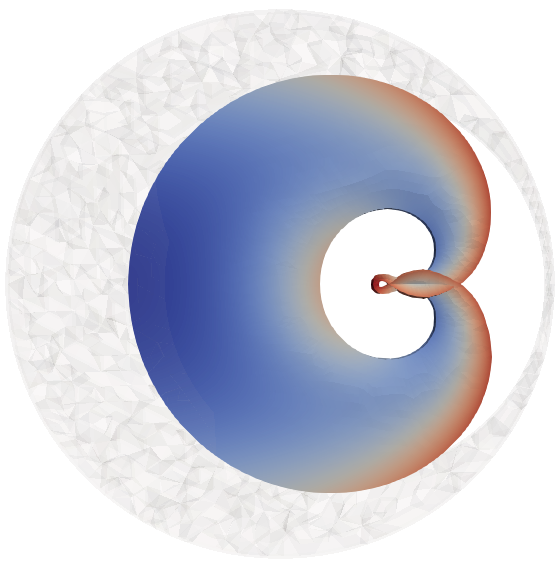}
        \caption{\SI{0.5e5}{\pascal}}
    \end{subfigure}
    \caption{Solution of the inverse displacement problem on the eclipse geometry. The stress-free configuration is computed for (a) \SI{0.25e5}{\pascal} and (b) \SI{0.5e5}{\pascal}.}
    \label{fig:halfmoon-solution}
\end{figure}

We first load the semi-circle geometry with an endocardial pressure of \SI{400}{\pascal}, and we solve under these conditions the inverse displacement problem \eqref{eq:backward weak}. The solution is displayed in Figure~\ref{fig:semicircle-solution}, where the thicker part of the geometry virtually does not deform, and indeed all deformation is obtained from the thinner part of the geometry. This results in an endocardial interpenetration. We proceed analogously with the eclipse, where we depict the solution in Figure~\ref{fig:halfmoon-solution}. In contrast to the semi-circle case, here we see that there is a self-intersection through the epicardium.

These two examples of self-intersection represent a global geometric incompatibilities as detailed in Section~\ref{sec:self-penetration}, but they are still the solution obtained through the inverse displacement problem \eqref{eq:backward weak}. This means that, unless a contact formulation is used, there is no guarantee that the stress-free configuration will avoid self-penetrations. Furthermore, it is not trivial to formulate a contact inverse displacement problem that is compatible with the forward problem.

\section{Algorithms for solving the inverse problem}\label{section:algorithms}
There are essentially two approaches for solving \eqref{eq:backward weak}. The first one is to solve the weak formulation associated with the inverse problem \eqref{eq:backward weak}, e.g. by the Newton-Raphson method, and the second one is to leverage only \eqref{eq:forward weak}, known as the Sellier method. The target problem is computationally challenging in both formulations, so we use a simple homotopy strategy to increase the loading terms with a fixed step size, i.e. by ramping the loads. We will denote this operation with a pseudo-time parameter, such that a load $\vec f$ becomes $\vec f(t)=t \vec f$, with $t$ in $[0,1]$ being the ramp parameter. The desired solution is obtained when $t=1$. We show how to solve the inverse displacement problem with this strategy in Algorithm~\ref{alg:inv}, where the solution of problem \eqref{eq:backward weak} is done with a Newton algorithm. Unless stated otherwise, all nonlinear algorithms consider as initial guess the solution at the previous step.

\begin{algorithm}[!ht]
    \caption{Inverse displacement method with load ramp.}
    \label{alg:inv}
    \begin{algorithmic}[1]
        \STATE {\bf Input:} Initial point $\vec u^0$, $N$
        \STATE Set $\texttt{error}=1$, $\texttt{k} = 0$, $\Delta t = 1/N$, $t = \Delta t$, $\vec u^k = \vec u^0$
        \WHILE{$t \leq 1$}
        \STATE Compute solution $\vec d^k$ of \eqref{eq:backward weak} at pseudo-instant $t$.
        \STATE $t \leftarrow t + \Delta t$
        \ENDWHILE
        \RETURN{Solution $\vec d = \vec d^k$}
    \end{algorithmic}
\end{algorithm}

The most widely used method to compute the solution of \eqref{eq:backward weak} is known as the Sellier method~\cite{sellier2011iterative}. If we consider a relaxation parameter $\alpha>0$ and an initial displacement $\vec u^{(0)}$, Denoting by $\Omega(\vec X^k)$ the configuration obtained in the $\vec X^k$ coordinates, the algorithm is displayed in Algorithm~\ref{alg:sellier}. This method is a fixed-point iteration, which are in general prone to instabilities and lack of convergence. This has been alleviated by including an acceleration technique known as Aitken acceleration~\cite{genet2015heterogeneous,rausch2017augmented}, and further improved by an Armijo line search strategy~\cite{marx2022robust}. We will refer to the latter as the Aitken-Armijo strategy. The resulting method enjoys improved robustness, which makes it more reliable for data intensive applications. Still, it has been observed that Anderson acceleration performs better than Aitken acceleration in most practical applications (see for example Refs.~\cite{bucelli2022partitioned},~\cite{deng2023fast}). This can be explained mainly by two things: on one hand, Anderson acceleration can be regarded as a nonlinear variant of the GMRES algorithm, so it has better mathematical foundations. On the other hand, it uses an arbitrary number of previous iterations, whereas Aitken uses only one previous solution. We propose a single algorithm that can be used to choose between the Armijo-Aitken strategy and Anderson acceleration in Algorithm~\ref{alg:sellier-augmented}, where we have observed that combining both Aitken and Anderson never yields a better solver (not reported). One possible explanation for this is that Anderson is not capable of accelerating arbitrary fixed point iterations. Indeed, it has been shown that it can accelerate  linearly converging sequences, that quadratically convergent sequences may worsen their performance and anything in between is still an open problem~\cite{evans2020proof}.

Convergence of the Sellier method (all the three variants previously shown) is established when the deformed geometry is sufficiently close to the original one, or when the increments are sufficiently small. The latter can lead to stagnation, which we have observed to happen sometimes with Aitken acceleration. For this, we have set a minimum relaxation of 0.5 that truncates smaller values.

\begin{algorithm}[!ht]
    \caption{Sellier method with load ramp.}
    \label{alg:sellier}
    \begin{algorithmic}[1]
        \STATE {\bf Input:} Initial point $\vec u^0$, relaxation $\alpha>0$, tolerance \texttt{tol}, and maximum iterations \texttt{maxit}
        \STATE Set $\texttt{error}=1$, $\texttt{k} = 0$, $\vec u^k = \vec u^0$
        \WHILE{$t<1$}
        \WHILE{$\texttt{error}>\texttt{tol}$ and $\texttt{k}<\texttt{maxit}$}
        \STATE Compute forward displacement $\vec d^k$  at $t$ in $\Omega(\vec X^k)$

        \STATE Compute the incremental displacement $\vec \delta^k = \vec X^k + \vec d^k - \vec X^0$
        \STATE Deform the geometry with displacement $-\vec \delta^k$
        \STATE Update $\texttt{error}$, $k \leftarrow k+1$
        \ENDWHILE
        \STATE $t \leftarrow t + \Delta t$
        \ENDWHILE
        \RETURN{Solution $\vec d = \vec d^k$}
    \end{algorithmic}

\end{algorithm}

\begin{algorithm}[!ht]
    \caption{Generalized Sellier method with load ramp.}
    \label{alg:sellier-augmented}
    \begin{algorithmic}[1]
        \STATE {\bf Input:} Initial point $\vec u^0$, relaxation $\alpha>0$, tolerance \texttt{tol}, and maximum iterations \texttt{maxit}
        \STATE Set $\texttt{error}=1$, $\texttt{k} = 0$, $\vec u^k = \vec u^0$
        \WHILE{$t\leq1$}
        \WHILE{$\texttt{error}>\texttt{tol}$ and $\texttt{k}<\texttt{maxit}$}
        \IF{Do Armijo:}
        \FOR{$\ell$ in $\{1,1/2, \hdots, \ell_\text{min}\}$}
        \STATE Compute forward displacement $\vec d^\ell$ at $t$ in $\Omega(\vec X^k)$
        \STATE Compute increment $\vec \delta^\ell = \vec X^\ell + \vec d^\ell- \vec X^0$
        \STATE If $k>0$, set $\alpha^\ell = -\alpha \frac{<\vec \delta^{k-1}, \vec \delta^\ell - \vec \delta^{k-1}>}{|\vec \delta^\ell - \vec\delta^{k-1}|^2}$
        \STATE Compute $\vec X^\ell = \vec X^k - \ell \alpha^\ell \vec \delta^\ell$
        \STATE Compute line search error $\texttt{error}_\ell$
        \STATE {If $k=0$ or $\texttt{error}_\ell < \texttt{error}$: \textbf{break}}
        \ENDFOR{ // for $\ell$}
        \STATE Find $\ell^*$ with minimum $\texttt{error}_\ell$
        \STATE Update displacement $\vec d^k = \vec d^{\ell^*}$ and $\alpha=\alpha^\ell$
        \ELSE
        \STATE Compute forward displacement $\vec d^k$  at $t$ in $\Omega(\vec X^K)$
        \ENDIF{ // Do Armijo}

        \IF{Do Anderson:}
        \STATE Compute accelerated solution $\vec d^k = AA(\vec d^k,\hdots, \vec d^{k-m})$
        \ENDIF{ // Do Anderson}
        \STATE Compute the incremental displacement $\vec \delta^k = \vec X^k + \vec d^k - \vec X^0$
        \STATE Deform the geometry with displacement $-\alpha \vec \delta^k$
        \STATE Update $\texttt{error}$, $k \leftarrow k +1$
        \ENDWHILE
        \STATE $t \leftarrow t + \Delta t$
        \ENDWHILE
        \RETURN{Solution $\vec d = \vec d^k$}
    \end{algorithmic}
\end{algorithm}

\section{Numerical tests}
\label{section:numerical-tests}
We perform the numerical tests in four different geometries:
\begin{enumerate}
    \item A 2D square domain.
    \item A 3D rectangular geometry, commonly referred to as \textit{slab} in the computational cardiology community, subject to surface and volume loads to validate our solvers.
    \item A simplified left ventricle (LV) geometry subject to an endocardial pressure and an active stress force.
    \item A realistic full-heart geometry with given physiological values of atrial and ventricular pressures.
\end{enumerate}

The scope of this section is to clarify the following points: (i)  to understand whether it is advantageous to solve the IEP using the Cauchy formulation or the Eshelby one, (ii) to compare the performance of IEP by using its direct solution or a Sellier approach in terms of its robustness (behavior with varying parameters) and optimality (sensitivity on problem size), and (iii) to characterize the computational effort of the IEP with respect to the DEP, i.e.\, which problem is most computationally challenging, and how to measure this aspect. For these aims, the numerical tests we propose are the following.

\begin{enumerate}
    \item A numerical convergence test for the Cauchy and Eshelby formulations for varying degrees of approximation. This test will help us conclude which of the two formulations should be used in practice.
    \item A robustness test where we vary the load of the slab and the endocardial pressure/active stress of the idealized LV. This test measures the sensitivity of the solvers with respect to external loads.
    \item An optimality test in which, for fixed loads, we increase the degrees of freedom of each problem. This test measures the sensitivity of the solvers with respect to the problem size.
    \item A preconditioning test, where we study the performance of both algebraic multigrid (AMG) and domain decomposition (DD) methods for the IEP formulation.
    \item A formulation comparison test, in which we study whether the backward or forward problems are more computationally demanding.
    \item A real-case scenario where we can test our conclusions in a full-heart model.
\end{enumerate}

In what follows, we will use the term \textit{inverse displacement method} to denote a direct numerical approximation of the IEP, based either on a finite element approximation of the Cauchy version \eqref{eq:backward weak} or the Eshelby one \eqref{eq:eshelby_weak}.
The inverse displacement method is thus a way, alternative to the Sellier's method, to solve the IEP, and should not be confused with the latter.

To avoid ambiguity, we will consider the nonlinear iterations to be the number of iterations required for each method to converge. For the inverse displacement method, this will be the number of Newton iterations. Instead, for the Sellier method, this will refer to the fixed point iterations required for convergence. Given that at each fixed point iteration this method incurs on the solution of a nonlinear elasticity problem, we will refer to such iterations as the inner nonlinear iterations. Whenever more than one ramp step is used, we will report the average number of iterations. The implementation of all tests on the slab and on the idealized LV have been implemented with the FEniCS library~\cite{alnaes2015fenics} and visualized with Paraview~\cite{henderson2007view}. The preconditioning tests have been performed with the Firedrake library~\cite{rathgeber2016firedrake}.
In addition, unless stated otherwise, all linear systems are solved using the MUMPS library~\cite{amestoy2000mumps}, which uses a direct method. This avoids the additional complexity of considering the challenges associated with the linear system resolution whenever quantifying the computational burden of the IEP.
The real-case scenario was performed with the high-performance \texttt{c++} library \texttt{life\textsuperscript{x}} (see\footnote{\url{https://lifex.gitlab.io/}} and~\cite{africa2022lifex}), built upon the finite element core \texttt{deal.II} (see\footnote{\url{https://www.dealii.org}} and~\cite{dealII91}).

\subsection{Numerical convergence test}\label{sec:convergence}

In this section, we propose a simple convergence analysis of the discretized counterparts of the weak formulations \eqref{eq:backward weak}-\eqref{eq:eshelby_weak}. We consider the 2D square domain $\Omega=[0,\,L]\times[0,\,L]$, and assume that the body is homogeneous and composed of a material with Neo-Hookean strain energy \eqref{eq:neo-hookean}. We construct the fields $\vec b$ and $\vec t$ such that
\begin{equation}
    \label{eq:usin}
    \widehat{\vec u}(\vec{x}) = A \sin(2 \pi x_1)\vec{e}_2
\end{equation}
is a solution of the inverse problem. In \eqref{eq:usin}, $\vec x = x_1 \vec{e}_1 + x_2 \vec{e}_2$ and $(\vec{e}_1,\,\vec{e}_2)$ is the canonical basis in $\mathbb{R}^2$.

We can now compute the corresponding Cauchy stress tensor through \eqref{eq:Cauchy_NH} and, by applying \eqref{eq:balance_linear_cauchy} we can recover the corresponding fields $\vec b$ and $\vec t$. Similarly, from \eqref{eq:betasigma} we can recover the expressions of $\vec \beta$ and $\vec \sigma$ such that \eqref{eq:usin} is a solution of \eqref{eq:balance_eshelby}.

We use this analytical solution to perform the convergence analysis of the discretized problem. We exploit a Galerkin approximation and the finite element method. We construct a triangular, structured mesh $\Omega_h$ of the domain $\Omega$, with $h$ being the diagonal of the elements. We use $P^1,\,P^2$ and $P^3$ elements to discretize the field $\widehat{\vec u}$ and we denote by $\widehat{\vec u}_h$ the discrete counterpart. The nonlinear problem is solved by means of a Newton method.

\begin{figure}[ht!]
    \centering
    \includegraphics[width=0.5\textwidth]{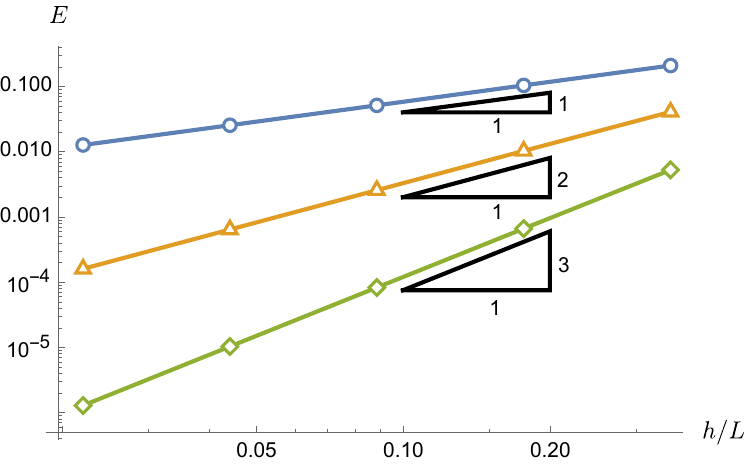}%
    \includegraphics[width=0.5\textwidth]{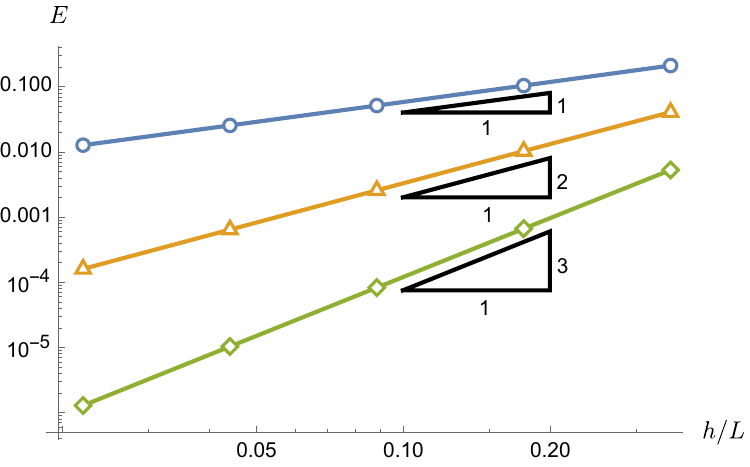}
    \caption{Results of the convergence analysis for the problem described in Section~\ref{sec:convergence}. Here we show $E$, i.e. the error norm $\|\widehat{\vec{u}}-\widehat{\vec{u}}_h\|_{H^1}$ non-dimensionalized with respect to the length-scale $L$, as a function of $h/L$ for the weak formulations \eqref{eq:backward weak} (left) and \eqref{eq:eshelby_weak} (right) for $\lambda/\mu=1000$. The blue, orange, and green markers correspond to the solutions obtained with $P^1$, $P^2$, and $P^3$ elements, respectively. The two plots are almost identical and we do not have significant differences in the marker positions.}
    \label{fig:convergence}
\end{figure}
In Figure~\ref{fig:convergence}, we show a logarithmic plot of the error norm $\|\widehat{\vec{u}}-\widehat{\vec{u}}_h\|_{H^1(\Omega,\,\mathbb{R}^2)}$ for $A = 0.1\,L$. We observe that the error is $O(h^n)$ for the element $P^n$ as $h\rightarrow 0$. The errors measured using the weak forms \eqref{eq:backward weak} and \eqref{eq:eshelby_weak} are very close, even though the formulation using the Eshelby stress \eqref{eq:eshelby_weak} requires much more iterations. Indeed, for the formulation with the Cauchy stress tensor, we can solve the problem with a single Newton algorithm which requires an average of $5.2$ inner iterations. Conversely, with the Eshelby stress the direct application of the Newton method may fail and we need to use a ramp where we iteratively increase the value of $A$, see Table~\ref{tab:convergence}.  Therefore, in the remaining part of this work, we will focus our attention on the Cauchy stress weak form \eqref{eq:backward weak}.

\begin{table}[]
    \centering
    \begin{tabular}{@{}cc|c|cc@{}}
        \toprule
            &         & \textit{Cauchy formulation} & \multicolumn{2}{c}{\textit{Eshelby formulation}}                     \\\midrule
        $n$ & Element & Newton iterations           & Ramp steps                                       & Newton iterations \\\midrule
        4   & P1      & 5                           & 2                                                & 6.5               \\
        8   & P1      & 5                           & 2                                                & 7.0               \\
        16  & P1      & 5                           & 2                                                & 10.0              \\
        32  & P1      & 5                           & 9                                                & 5.3               \\
        64  & P1      & 5                           & 10                                               & 5.7               \\\midrule
        4   & P2      & 5                           & 3                                                & 8.0               \\
        8   & P2      & 5                           & 4                                                & 9.0               \\
        16  & P2      & 5                           & 7                                                & 7.1               \\
        32  & P2      & 5                           & 11                                               & 5.7               \\
        64  & P2      & 6                           & 12                                               & 5.7               \\\midrule
        4   & P3      & 5                           & 4                                                & 7.8               \\
        8   & P3      & 5                           & 10                                               & 5.5               \\
        16  & P3      & 5                           & 11                                               & 5.6               \\
        32  & P3      & 6                           & 12                                               & 5.7               \\
        64  & P3      & 6                           & 13                                               & 5.6               \\\bottomrule
    \end{tabular}
    \caption{Average number of Newton iterations per ramp step and ramp steps necessary for solving the problem described in Section~\ref{sec:convergence}. Here, we use just one ramp step for the Cauchy formulation and $n$ is the number of element for each side of the square domain.}
    \label{tab:convergence}
\end{table}

\subsection{Slab tests}\label{sec:beam-tests}
\begin{figure}[ht!]
    \centering
    \begin{subfigure}[b]{0.49\textwidth}
        \includegraphics[width=\textwidth]{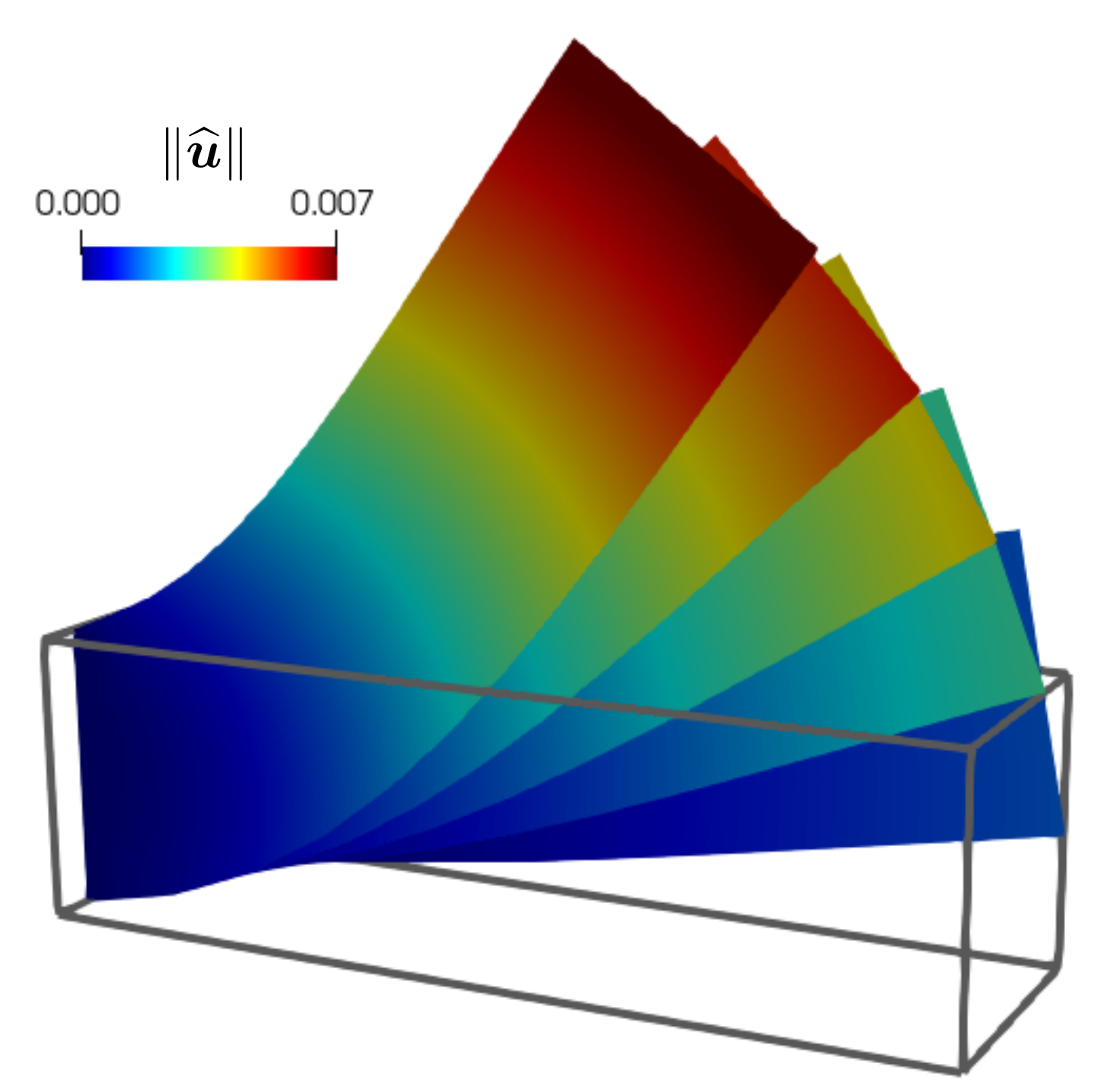}
        \caption{Varying volumetric load.}
    \end{subfigure}
    \begin{subfigure}[b]{0.49\textwidth}
        \includegraphics[width=\textwidth]{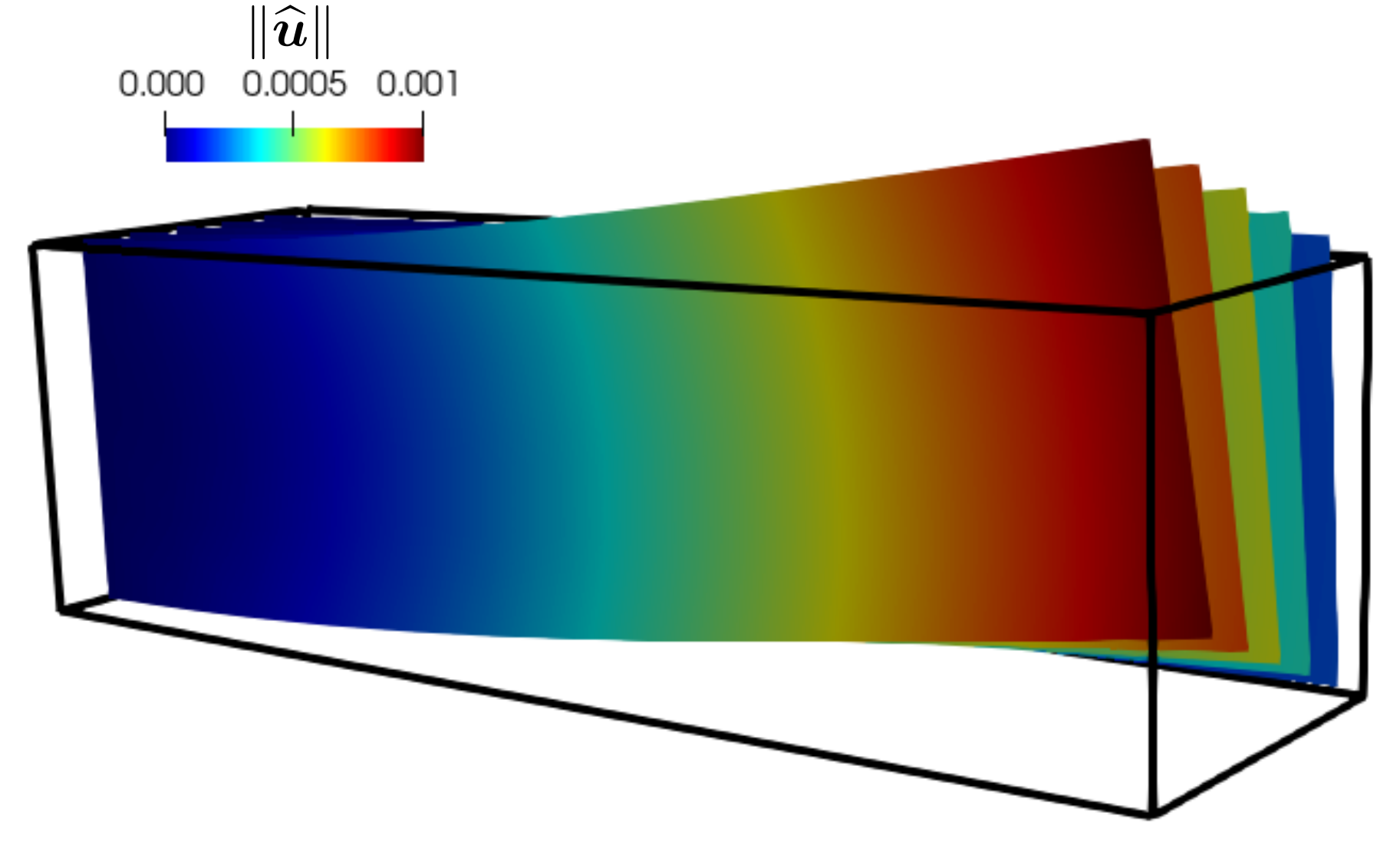}
        \caption{Varying surface load.}
    \end{subfigure}
    \caption{Solution of the inverse displacement problem computed on a slab. In Figure (a), we compute the stress-free configuration for forces $\vec b=-b\vec{\hat{e}}_z$ with $b\in\{\SI{10}{\pascal\per\meter},\SI{20}{\pascal\per\meter},\SI{30}{\pascal\per\meter},\SI{40}{\pascal\per\meter},\SI{50}{\pascal\per\meter}\}$ with surface load $\vec t=\vec 0$. In Figure (b) we do the same for $\vec b=\vec 0$ and $\vec t=-t\vec{\hat{e}}_z$ with $t\in\{\SI{5}{\pascal},\SI{10}{\pascal},\SI{15}{\pascal},\SI{20}{\pascal},\SI{25}{\pascal}\}$.}
    \label{fig:slab-id-solution}
\end{figure}

The slab consists of a prism cut going from the endocardium to the epicardium, given by
\[
    \Omega \coloneqq (0,\,\SI{e-2}{\meter})\times (0,\,\SI{3e-3}{\meter})\times(0,\,\SI{3e-3}{\meter}).
\]
On it, we consider the exponential constitutive law of Usyk~\cite{Usyk2002}, detailed in Section~\ref{sec:cardio_results}, with homogeneous Dirichlet conditions on $\{x=0\}$ and null traction conditions elsewhere. We display the solution of the inverse displacement problem in Figure \ref{fig:slab-id-solution}, which we computed for various volumetric and surface loads, given by  $\vec b$ and $\vec t$ respectively in \eqref{eq:balance_linear_cauchy}. We note that both solutions were computed using 10 ramp steps for the loads, and the maximum load used for each display was such that twice bigger loads would yield a divergent iterative procedure using the IEP formulation.

\subsubsection{Robustness}
We study the robustness with respect to volumetric loads. To measure the performance, we look at the number of nonlinear iterations required for convergence. All tests were performed with a Newton method using absolute and relative tolerances of $10^{-14}$ and $10^{-6}$ respectively for the inverse problem. The Sellier methods use equal absolute and relative tolerances of $10^{-6}$. The tangent systems were inverted with MUMPS, a parallel direct solver. The geometry was discretized with 24 subdivisions in the $x$ direction, and 8 subdivisions in the $y$ and $z$ directions, resulting in roughly $6\,000$ degrees of freedom.

We show the results of this test in Table~\ref{table:robustness-slab}. We first note that the inverse displacement method is much more robust than the Sellier methods in general, being able to yield a solution for load values more roughly 10 times larger than those of Sellier methods, in only one ramp step, and 4 times larger if Sellier uses 100 ramp steps. Among Sellier methods, we note that they all converge in the same scenarios, meaning that acceleration does not make a difference in this test. Still, it can be appreciated how the Armijo strategy yields a more robust method, which can be greatly improved by using instead Anderson acceleration. Indeed, the latter can sometimes yield convergence in roughly half the number of nonlinear iterations. Still, the superiority of this strategy is less obvious when looking at the inner nonlinear iterations, which increase as the accelerated methods perform larger steps. Naturally, this nested solver problem is not present in the inverse displacement method.

\begin{table}
    \centering
    \newcommand{\mc}[1]{\multicolumn{3}{|c|}{#1}}
    \begin{footnotesize}
        \begin{subtable}{\textwidth}
            \centering
            \begin{tabular}{|r|rrr|rrr|}
                \toprule Vol load & \mc{ID} & \mc{Sellier}                                            \\ \midrule
                                  & 1       & 10           & 100 & 1         & 10         & 100       \\ \midrule
                1                 & 3       & 2            & 2   & 2.0 (2.3) & 1.0 (2.0)  & 1.0 (1.9) \\
                10                & 4       & 3            & 2   & 8.0 (3.3) & 3.7 (2.5)  & 2.0 (2.0) \\
                25                & 5       & 3            & 3   & --        & 17.0 (3.0) & 8.0 (2.5) \\
                50                & 6       & 4            & 3   & --        & --         & --        \\
                75                & 7       & 4            & 3   & --        & --         & --        \\
                100               & --      & 4.1          & 3   & --        & --         & --        \\ \bottomrule
            \end{tabular}
        \end{subtable}

        \begin{subtable}{\textwidth}
            \centering
            \begin{tabular}{|r|rrr|rrr|}
                \toprule Vol load & \mc{Sellier Armijo} & \mc{Sellier Anderson}                                                 \\ \midrule
                                  & 1                   & 10                    & 100       & 1         & 10        & 100       \\ \midrule
                1                 & 2.0 (2.3)           & 1.0 (2.0)             & 1.0 (1.9) & 2.0 (2.3) & 1.0 (2.0) & 1.0 (1.9) \\
                10                & 9.0 (3.2)           & 3.9 (2.5)             & 2.0 (2.0) & 4.0 (4.0) & 3.2 (2.5) & 2.0 (2.0) \\
                25                & --                  & 12.0 (4.5)            & 6.5 (3.6) & --        & 4.7 (3.2) & 3.5 (2.6) \\
                50                & --                  & --                    & --        & --        & --        & --        \\
                75                & --                  & --                    & --        & --        & --        & --        \\
                100               & --                  & --                    & --        & --        & --        & --        \\ \bottomrule
            \end{tabular}
        \end{subtable}
    \end{footnotesize}

    \caption{Slab robustness. Average number of nonlinear iterations (average inner nonlinear iterations in parenthesis) when increasing the volumetric load with 1 ramp step, 10 ramp steps and 100 ramp steps. ID: Inverse displacement, and Anderson acceleration is used with a depth of 3.}
    \label{table:robustness-slab}
\end{table}

\subsubsection{Optimality}
In this section, we study the sensitivity of the slab problem as the number of degrees of freedom increases. For this, we consider two volumetric loads given by $\vec b=-b\vec{e}_3$ for $b$ in $\{\SI{10}{\pascal\per\meter},\SI{20}{\pascal\per\meter}\}$, and we divide the $x$, $y$, and $z$ axes into $3k$, $k$, and $k$ elements respectively, for $k$ in $\{2,4,8,16,24,32,40\}$, solved in one ramp step. We show the results in Table~\ref{table:optimality-slab}, where we highlight the following results: (i) as typical in Newton methods, the IEP formulation behaves optimally, with its number of nonlinear iterations remaining constant as the number of degrees of freedom increase~\cite{kelley1991mesh}. (ii) For the smaller load test, the pure Sellier and its Armijo variant are vastly more reliable than the Anderson accelerated variant. Still, for larger loads they all behave very erratically, and there is no obvious better option.

\begin{table}
    \centering
    \begin{subtable}{\textwidth}
        \centering
        \begin{tabular}{|r|r|c|c|c|c|}
            \toprule $N$ & DoFs   & ID  & Sellier   & Sellier Armijo & Sellier Anderson \\ \midrule
            2            & 189    & 4.0 & 5.0 (3.0) & 5.0 (3.0)      & 3.0 (3.5)        \\
            4            & 975    & 4.0 & 6.0 (3.1) & 7.0 (2.9)      & 4.0 (3.4)        \\
            8            & 6075   & 4.0 & 8.0 (3.3) & 9.0 (3.2)      & 4.0 (4.0)        \\
            16           & 42483  & 4.0 & 8.0 (3.2) & 9.0 (3.2)      & --               \\
            24           & 136875 & 4.0 & 8.0 (3.2) & 9.0 (3.1)      & 4.0 (4.8)        \\
            32           & 316899 & 4.0 & 8.0 (3.2) & 9.0 (3.2)      & --               \\
            40           & 610203 & 4.0 & 7.0 (3.4) & 8.0 (3.3)      & --               \\ \bottomrule
        \end{tabular}
        \caption{Results with $\vec b=-10\vec{e}_3$.}
    \end{subtable}

    \begin{subtable}{\textwidth}
        \centering
        \begin{tabular}{|r|r|c|c|c|c|}
            \toprule $N$ & DoFs   & ID  & Sellier    & Sellier Armijo & Sellier Anderson \\ \midrule
            2            & 189    & 4.0 & --         & 10.0 (3.1)     & 5.0 (3.8)        \\
            4            & 975    & 4.0 & 17.0 (3.6) & --             & 7.0 (3.5)        \\
            8            & 6075   & 5.0 & --         & --             & 8.0 (4.9)        \\
            16           & 42483  & 5.0 & 31.0 (3.8) & 25.0 (5.9)     & --               \\
            24           & 136875 & 5.0 & --         & --             & --               \\
            32           & 316899 & 5.0 & --         & --             & --               \\
            40           & 610203 & 5.0 & --         & --             & --               \\ \bottomrule
        \end{tabular}
        \caption{Results with $\vec b=-20\vec{e}_3$.}
    \end{subtable}
    \caption{Slab optimality. Average number of nonlinear iterations (and inner nonlinear iterstions) when increasing the number of degrees of freedom. (DoFs) Degrees of freedom, (ID) inverse displacement.}
    \label{table:optimality-slab}
\end{table}

\subsubsection{Performance comparison}
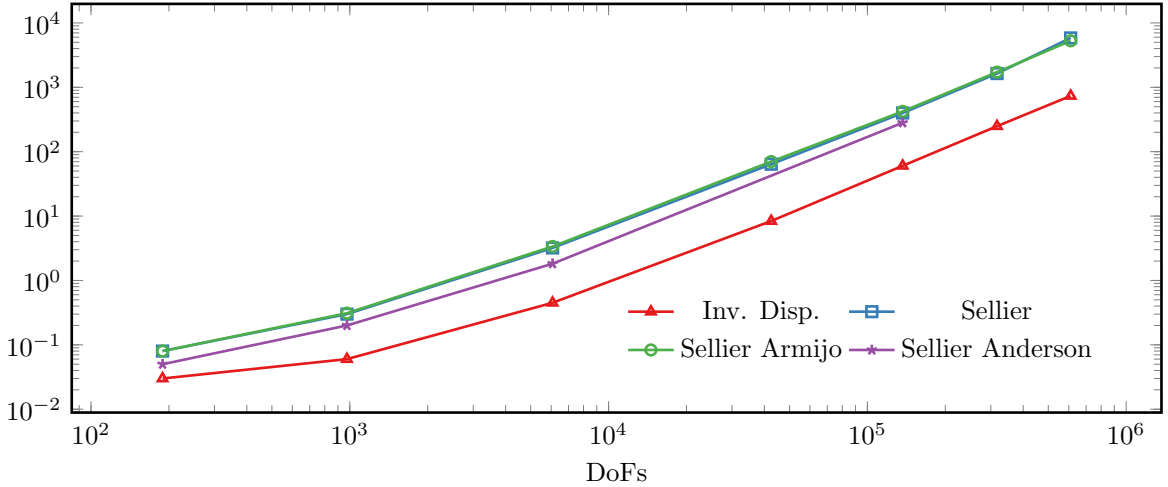
\begin{figure}
    \centering

    \begin{tikzpicture}
    \begin{loglogaxis}
        [width=\textwidth, height=7cm, xlabel=DoFs, ylabel=, tick label style={font=\normalsize},line width=1.0pt,legend style={draw=none, fill opacity=1.0, text opacity= 1,row sep =2pt, font=\normalsize, at={(0.95,0.2)}, anchor=east}, legend columns=2, cycle list/Set1-5]
        \addplot+[mark=triangle] coordinates {
                                                (189,    0.03)
                                                (975,    0.06)
                                                (6075,   0.45)
                                                (42483,  8.37)
                                                (136875, 60.36)
                                                (316899, 248.85)
                                                (610203, 735.73)
                                                };
        \addplot+[mark=square] coordinates {
                                                (189,    0.08)
                                                (975,    0.30)
                                                (6075,   3.20)
                                                (42483,  63.89)
                                                (136875, 398.87)
                                                (316899, 1636.14)
                                                (610203, 5844.36)
                                                };
        \addplot+[mark=o] coordinates {
                                                (189,    0.08)
                                                (975,    0.31)
                                                (6075,   3.37)
                                                (42483,  70.20)
                                                (136875, 422.18)
                                                (316899, 1720.43)
                                                (610203, 5275.55)
                                                };
        \addplot+[mark=star] coordinates {
                                                (189,    0.05)
                                                (975,    0.20)
                                                (6075,   1.82)
                                                (42483,  0)
                                                (136875, 281.60)
                                                (316899, 0)
                                                (610203, 0)
                                                };
        \legend{Inv. Disp., Sellier, Sellier Armijo, Sellier Anderson}
    \end{loglogaxis}
    \end{tikzpicture}

    \caption{Slab performance. Total CPU time employed to solve the problem for each method under consideration. (DoFs) Degrees of freedom, (ID) inverse displacement. Missing points mean that the procedure did not converge.}
    \label{table:performance-slab}
\end{figure}

In this section we compare the CPU times (also referred to as walltime) of the methods under consideration. For this, we present them for the first scenario considered in the optimality test, i.e. for the load $\vec b=-\SI{10}{\pascal\per\meter}\vec{e}_3$, and report them in Figure~\ref{table:performance-slab}. We note that the inverse displacement method provides and clear improvement over the Sellier method, representing roughly a speed-up of an 87\%. We highlight that, whenever the Anderson method converges, it is faster than both Sellier variants present in literature. Additionally, we confirm the overall superiority of the Armijo line search strategy for this case, as it is both more robust and faster than plain Sellier.

\subsubsection{Computational effort of IEP and DEP}\label{section:difficulty-slab}
For measuring which formulation is the most challenging at the numerical level, we use as an indicator the number of nonlinear iterations incurred by the nonlinear solver, if it converges. We do so in three scenarios: (i) the IEP formulation \eqref{eq:backward weak}, (ii) the DEP formulation from the stress-free configuration \eqref{eq:forward weak}, and (iii) the DEP from the current configuration \eqref{eq:forward weak}. The distinction between the last two is important because they represent two conceptually different scenarios. In scenario (ii), we compare the inverse and forward problems in a physically consistent setting. In problem (iii), the scope is purely methodological, as we compare the computational effort of the inverse problem with respect to what is done by the Sellier method.
As a matter of fact, the fixed-point iterations of the Sellier method envisage a sequence of DEPs, moving from (iii) to (ii).
This should provide further evidence for the lack of convergence of the Sellier method, justified additionally by the requirement of solving a challenging nonlinear problem at each iteration. We fix the load to be $\vec b = -\SI{10}{\pascal\per\meter}\vec{e}_3$.

We show the results in Table~\ref{table:comparison-slab}. Albeit unintuitive, we note that the easiest problem is the inverse one, which has a consistently lower number of nonlinear iterations than the other two problems. Interestingly, the forward problem from the stress-free configuration in this problem is slightly harder than the one posed on the current geometry. We remark that, in this test case, we only focus on the nonlinear solver, and we disregard the challenges associated with the inner linear systems, as we are employing a direct linear solver. This aspect will be addressed later in Sections~\ref{section:preconditioning} and \ref{sec:cardio_results}.

\begin{table}
    \centering
    \begin{tabular}{|r|r|r|r|r|}
        \toprule Vol load & ID  & ID-forw & Forward \\ \midrule
        1                 & 2.0 & 2.0     & 2.0     \\
        10                & 3.0 & 3.0     & 3.0     \\
        25                & 3.0 & 3.9     & 3.9     \\
        50                & 4.0 & 4.0     & 4.0     \\
        75                & 4.0 & 4.9     & 4.0     \\
        100               & 4.1 & --      & 4.3     \\ \bottomrule
    \end{tabular}

    \caption{Computational effort study. The numbers stand for the average number of nonlinear iterations incurred by the Newton solver. (ID) Inverse displacement problem, (ID-forw) Forward problem from stress-free configuration, and (Forward) is the forward problem from the current configuration.}
    \label{table:comparison-slab}
\end{table}

\subsection{Simplified cardiac model}
In this section, we study as in Section \ref{sec:beam-tests} the robustness, optimality, performance, and computational effort of the inverse displacement formulation against standard and accelerated Sellier schemes on an idealized LV geometry. We consider the same physical model as in the slab test, with the only difference of having the physiologically motivated boundary condition of \eqref{eqn:BC_epicardium} on the epicardium.

\begin{figure}[ht!]
    \centering
    \begin{subfigure}[b]{0.3\textwidth}
        \includegraphics[width=\textwidth]{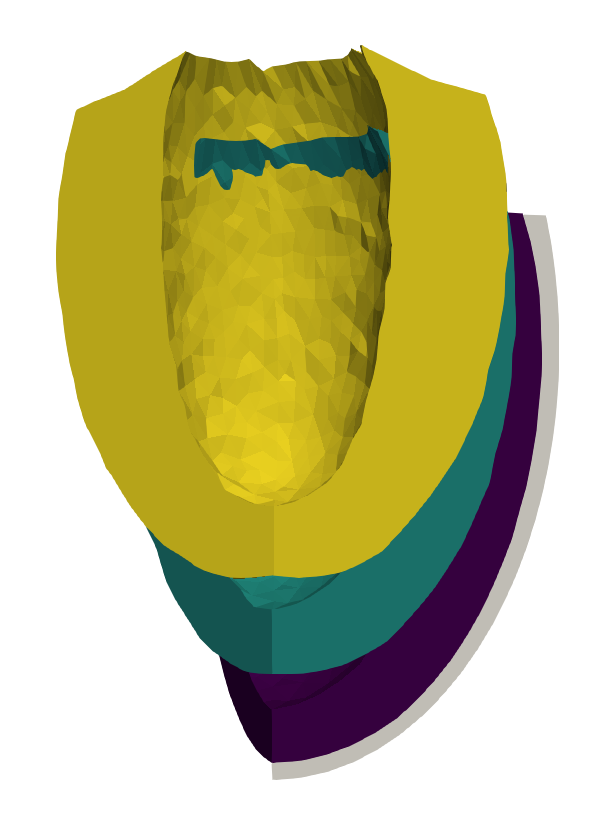}
    \end{subfigure}
    \begin{subfigure}[b]{0.3\textwidth}
        \includegraphics[width=\textwidth]{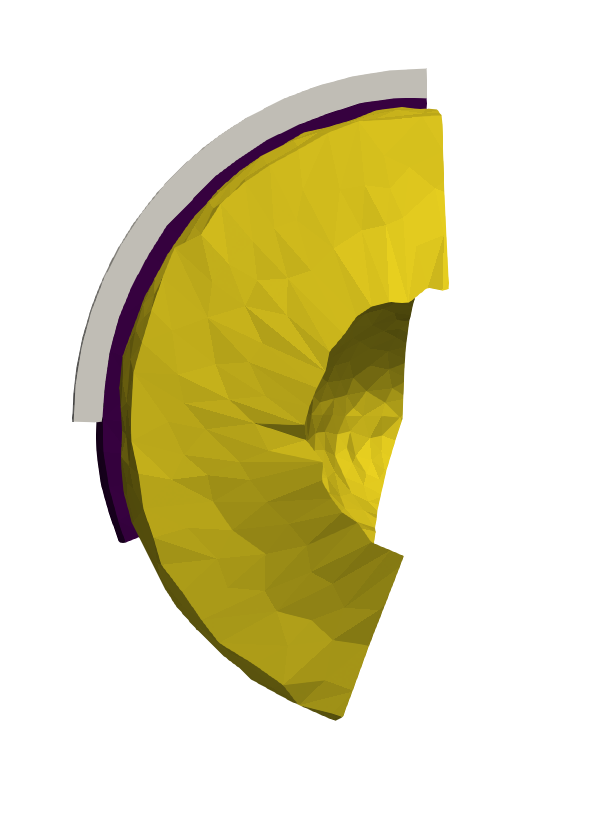}
    \end{subfigure}
    \caption{Stress-free configuration computed for pressures of $\SI{1}{\kilo\pascal}$ (blue), $\SI{5}{\kilo\pascal}$ (green), and $\SI{10}{\kilo\pascal}$ (yellow).}
    \label{fig:lv-id-solution-endo}
\end{figure}

\begin{figure}[ht!]
    \centering
    \includegraphics[height=4cm]{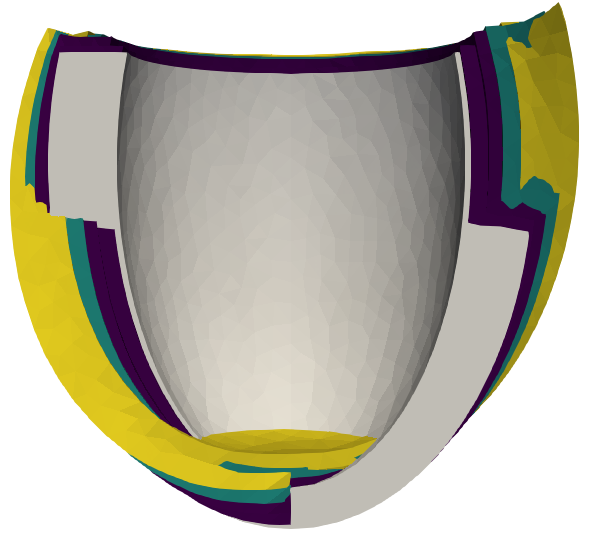}\quad
    \includegraphics[height=4cm]{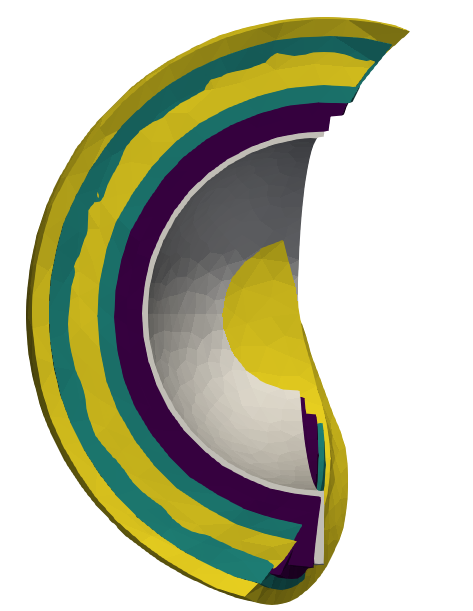}
    \caption{Stress-free configuration computed for active stress peaks given by $\SI{5}{\kilo\pascal}$ (blue), $\SI{10}{\kilo\pascal}$ (green), and $\SI{20}{\kilo\pascal}$ (yellow).}
    \label{fig:lv-id-solution-as}
\end{figure}

We focus on two types of loads, which are the main ones present in cardiac simulations: (i) a pressure acting uniformly on the endocardium and (ii) the active stress, which we depict respectively in Figures~\ref{fig:lv-id-solution-endo} and~\ref{fig:lv-id-solution-as} respectively.

\subsubsection{Robustness}
In this section we study the robustness of the methods with respect to an endocardial pressure going from $\SI{0.1}{\kilo\pascal}$ to $\SI{10}{\kilo\pascal}$ for 1, 10, and 100 ramp steps. Then, we do the same computation for an active stress magnitude given going from $\SI{1}{\kilo\pascal}$ up to $\SI{40}{\kilo\pascal}$.

The computed results are shown in Tables~\ref{table:robustness-lv-endo} and~\ref{table:robustness-lv-as} for the endocardial pressure and the active stress, respectively. First, we note that again the inverse displacement method is the most robust in all scenarios under consideration. There is no significant advantage in augmenting the standard Sellier method with the Armijo strategy, but instead Anderson acceleration provides a consistently more robust solver in both the number of nonlinear iterations and the scenarios in which it converges.  We also highlight that the inverse displacement formulation and the Anderson accelerated Sellier method yield the same robustness when using 100 ramp steps.

\begin{table}
    \centering
    \newcommand{\mc}[1]{\multicolumn{3}{|c|}{#1}}
    \begin{footnotesize}
        \begin{tabular}{|r|rrr|rrr|}
            \toprule Pa & \mc{ID} & \mc{Sellier}                                            \\ \midrule
                        & 1       & 10           & 100 & 1         & 10         & 100       \\ \midrule
            100         & 3.0     & 2.0          & 2.0 & 2.0 (2.3) & 1.0 (2.0)  & 1.0 (1.9) \\
            1000        & 8.0     & 3.3          & 2.4 & --        & 7.4 (2.6)  & 3.7 (2.1) \\
            2500        & --      & 4.2          & 3.0 & --        & 13.1 (2.7) & 6.5 (2.3) \\
            5000        & --      & 4.5          & 3.0 & --        & --         & 6.8 (2.3) \\
            10000       & --      & --           & 3.0 & --        & --         & --        \\ \bottomrule
        \end{tabular}

        \begin{tabular}{|r|rrr|rrr|}
            \toprule Pa & \mc{Sellier Armijo} & \mc{Sellier Anderson}                                                  \\ \midrule
                        & 1                   & 10                    & 100       & 1         & 10        & 100        \\ \midrule
            100         & 2.0 (2.3)           & 1.0 (2.0)             & 1.0 (1.9) & 2.0 (2.3) & 1.0 (2.0) & 1.0  (1.9) \\
            1000        & --                  & 7.2 (2.6)             & 3.6 (2.2) & --        & 4.4 (2.7) & 2.9  (2.2) \\
            2500        & --                  & 10.9 (2.8)            & 6.5 (2.4) & --        & 7.6 (2.8) & 4.8  (2.4) \\
            5000        & --                  & --                    & 6.7 (2.9) & --        & --        & 5.6  (2.3) \\
            10000       & --                  & --                    & --        & --        & --        & 6.6  (2.4) \\ \bottomrule
        \end{tabular}
    \end{footnotesize}

    \caption{Left ventricle robustness with respect to endocardial pressures. Average number of nonlinear iterations (average inner nonlinear iterations in parenthesis) when increasing the volumetric load with 1, 10, and 100 ramp steps. (ID) Inverse displacement, and Anderson acceleration is used with a depth of 3.}
    \label{table:robustness-lv-endo}
\end{table}

\begin{table}
    \centering
    \newcommand{\mc}[1]{\multicolumn{3}{|c|}{#1}}
    \begin{footnotesize}
        \begin{tabular}{|r|rrr|rrr|}
            \toprule Pa & \mc{ID} & \mc{Sellier}                                            \\ \midrule
                        & 1       & 10           & 100 & 1         & 10         & 100       \\ \midrule
            1000        & 4.0     & 3.0          & 2.0 & 3.0 (3.0) & 1.9 (2.3)  & 1.0 (2.0) \\
            5000        & --      & 3.2          & 2.5 & --        & 5.1 (2.6)  & 2.9 (2.1) \\
            10000       & --      & 3.4          & 2.6 & --        & 11.8 (2.6) & 6.4 (2.1) \\
            20000       & --      & 3.8          & 2.7 & --        & --         & --        \\
            40000       & --      & --           & 2.7 & --        & --         & --        \\ \bottomrule
        \end{tabular}

        \begin{tabular}{|r|rrr|rrr|}
            \toprule Pa & \mc{Sellier Armijo} & \mc{Sellier Anderson}                                                  \\ \midrule
                        & 1                   & 10                    & 100       & 1         & 10         & 100       \\ \midrule
            1000        & 4.0 (2.8)           & 1.9 (2.3)             & 1.0 (2.0) & 3.0 (3.0) & 1.9  (2.3) & 1.0 (2.0) \\
            5000        & --                  & 6.3 (3.8)             & 3.6 (2.8) & --        & 4.1  (2.6) & 2.7 (2.2) \\
            10000       & --                  & 6.9 (3.9)             & 4.4 (3.1) & --        & 5.5  (2.7) & 3.6 (2.2) \\
            20000       & --                  & --                    & --        & --        & --         & 4.6 (2.4) \\
            40000       & --                  & --                    & --        & --        & --         & 4.9 (2.4) \\ \bottomrule
        \end{tabular}
    \end{footnotesize}

    \caption{Left ventricle robustness with respect to the peak active stress. Average number of nonlinear iterations (average inner nonlinear iterations in parenthesis) when increasing the volumetric load with 1, 10, and 100 ramp steps. (ID) Inverse displacement, and Anderson acceleration is used with a depth of 3.}
    \label{table:robustness-lv-as}
\end{table}

\subsubsection{Optimality}
In this section we study the performance of all methods under consideration as the number of degrees of freedom increases. We consider two scenarios: one with a fixed endocardial pressure of $\SI{0.2}{\kilo\pascal}$ and another one with a fixed active stress peak of $\SI{5}{\kilo\pascal}$, with the results in Table~\ref{table:optimality-lv}. We note that in all considered scenarios, the inverse displacement method yields a more robust performance. Still, we highlight that Anderson acceleration performs roughly the same average inner nonlinear iterations as pure Sellier, with reduced nonlinear iterations. In the active stress case, the Armijo strategy is instead both more costly and less robust. Still, there is no significant difference among the methods tested.

\begin{table}
    \centering
    \begin{subtable}{\textwidth}
        \centering
        \begin{tabular}{|r|c|c|c|c|}
            \toprule DoFs & ID  & Sellier   & Sellier Armijo & Sellier Anderson \\ \midrule
            9375          & 3.0 & 1.4 (2.0) & 1.4 (2.0)      & 1.4 (2.0)        \\
            20709         & 3.0 & 1.3 (2.0) & 1.3 (2.0)      & 1.3 (2.0)        \\
            60081         & 3.0 & 1.1 (2.5) & 1.1 (2.5)      & 1.1 (2.5)        \\
            149511        & 3.1 & 2.2 (2.6) & 2.0 (3.0)      & 1.9 (2.6)        \\ \bottomrule
        \end{tabular}
        \caption{Endocardial pressure of $\SI{0.2}{\kilo\pascal}$.}
    \end{subtable}

    \begin{subtable}{\textwidth}
        \centering
        \begin{tabular}{|r|c|c|c|c|}
            \toprule DoFs & ID  & Sellier   & Sellier Armijo & Sellier Anderson \\ \midrule
            9375          & 3.4 & 5.1 (2.6) & 6.3 (3.8)      & 4.1 (2.6)        \\
            20709         & 3.4 & 4.5 (2.7) & 5.7 (3.6)      & 4.0 (2.7)        \\
            60081         & 3.4 & 5.5 (2.8) & --             & 4.0 (2.9)        \\
            149511        & 3.6 & 4.6 (3.1) & 5.9 (4.6)      & 3.9 (3.2)        \\ \bottomrule
        \end{tabular}
        \caption{Peak active stress of $\SI{5}{\kilo\pascal}$.}
    \end{subtable}

    \caption{LV optimality. We show the average nonlinear iterations (and average inner nonlinear iterations) in ten ramp steps, with an endocardial pressure of $\SI{0.2}{\kilo\pascal}$.}
    \label{table:optimality-lv}
\end{table}

\subsubsection{Performance comparison}
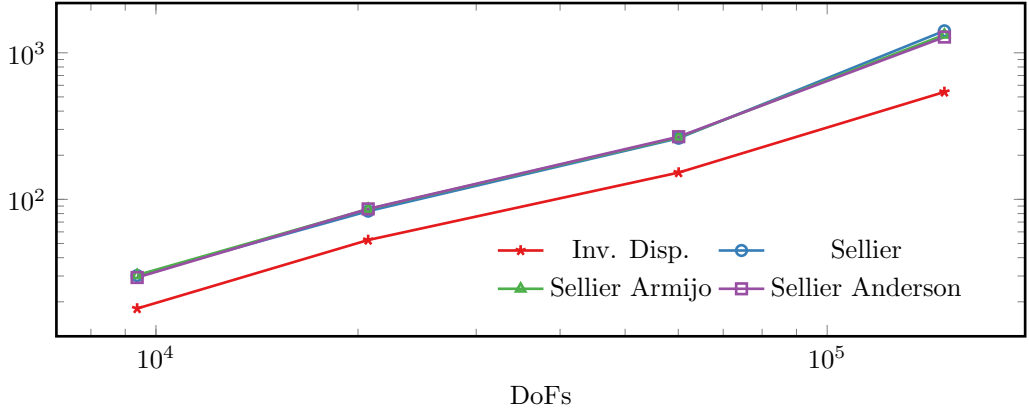
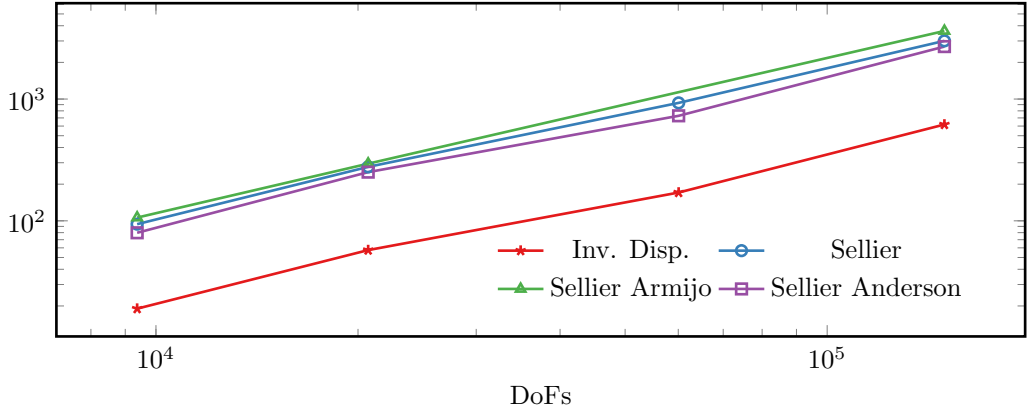
\begin{figure}
    \centering

    \begin{subfigure}{0.9\textwidth}
        \begin{tikzpicture}
        \begin{loglogaxis}
            [width=\textwidth, height=6cm, xlabel=DoFs, ylabel=, tick label style={font=\normalsize},line width=1.0pt,legend style={draw=none, fill opacity=0.7, text opacity= 1,row sep =2pt, font=\normalsize, at={(0.95,0.2)}, anchor=east}, legend columns=2, cycle list/Set1-5]
            \addplot+[mark=star] coordinates {
                                                    (9375,18.0)
                                                    (20709, 52.8)
                                                    (60081, 152.3)
                                                    (149511, 540.4)
                                                    };
            \addplot+[mark=o] coordinates {
                                                    (9375,30.2)
                                                    (20709, 83.1)
                                                    (60081, 261.3)
                                                    (149511, 1412.6)
                                                    };
            \addplot+[mark=triangle] coordinates {
                                                    (9375,30.4)
                                                    (20709, 85.8)
                                                    (60081, 264.6)
                                                    (149511, 1325.1)
                                                    };
            \addplot+[mark=square] coordinates {
                                                    (9375,29.3)
                                                    (20709, 86.1)
                                                    (60081, 267.6)
                                                    (149511, 1280.3)
                                                    };
            \legend{Inv. Disp., Sellier, Sellier Armijo, Sellier Anderson}
        \end{loglogaxis}
        \end{tikzpicture}
        \subcaption{Endocardial pressure}
    \end{subfigure}

    \begin{subfigure}{0.9\textwidth}
        \begin{tikzpicture}
        \begin{loglogaxis}
            [width=\textwidth, height=6cm, xlabel=DoFs, ylabel=, tick label style={font=\normalsize},line width=1.0pt,legend style={draw=none, fill opacity=0.7, text opacity= 1,row sep =2pt, font=\normalsize, at={(0.95,0.2)}, anchor=east}, legend columns=2, cycle list/Set1-5]
            \addplot+[mark=star] coordinates {
                                                    (9375,19.0)
                                                    (20709, 57.5)
                                                    (60081, 170.8)
                                                    (149511, 618.0)
                                                    };
            \addplot+[mark=o] coordinates {
                                                    (9375,93.9)
                                                    (20709, 277.4)
                                                    (60081, 930.1)
                                                    (149511, 3005.0)
                                                    };
            \addplot+[mark=triangle] coordinates {
                                                    (9375,106.3)
                                                    (20709, 294.7)
                                                    (60081, 0.0)
                                                    (149511, 3623.4)
                                                    };
            \addplot+[mark=square] coordinates {
                                                    (9375,79.8)
                                                    (20709, 251.1)
                                                    (60081, 728.5)
                                                    (149511, 2697.3)
                                                    };
            \legend{Inv. Disp., Sellier, Sellier Armijo, Sellier Anderson}
        \end{loglogaxis}
        \end{tikzpicture}
        \subcaption{Peak active stress}
    \end{subfigure}

    \caption{LV CPU times in seconds. (DoFs) Degrees of freedom, (ID) inverse displacement, and Anderson acceleration was used with a depth of 3. Missing points mean that the procedure did not converge.}
    \label{table:performance-lv}
\end{figure}

In this section, we compare the CPU times and report them in Figure~\ref{table:performance-lv}. In terms of execution time, we note that pure Sellier is the worst, and inverse displacement yields the best performance, yielding roughly a 60\% reduction in time with respect to pure Sellier. Between the two methods reside the Armijo and Anderson accelerated Sellier methods, which yield roughly a 7\% and a 10\% walltime reduction with respect to a pure Sellier method in the endocardial test. In the active stress case, Sellier Armijo is more expensive, whereas Anderson yields again a 10\% time save.

\subsubsection{Computational effort of the problem}\label{section:difficulty-lv}
In this section, we aim to study whether the inverse or forward problems are more computationally demanding as is Section~\ref{section:difficulty-slab} by considering the same three scenarios. In this case, as in the previous sections, we consider separately the effect of an endocardial pressure and that of an active stress force, shown in Table~\ref{table:comparison-lv}. We note that, as in the slab tests, the inverse displacement method requires the lowest number of iterations in almost all scenarios, except for some instances of active stress. Interestingly, in contrast to the slab case, this test shows that the forward problem from the computed stress-free configuration is easier than the one posed on the deformed configuration for all values considered. This is consistent with experience in cardiac modeling, and shows two things: on one hand, solving the problem from a stress-free configuration is easier, as it is more physically accurate for the given geometry. On the other hand, it shows why it is more difficult to get the Sellier method to converge. This suggest that the Sellier method could be made more robust by adding further ramping strategies to the inner nonlinear problem, which would result in even larger computational costs.

\begin{table}
    \centering
    \begin{subtable}{0.46\textwidth}
        \centering
        \begin{tabular}{|r|r|r|r|r|}
            \toprule Pa & ID  & ID-forw & Forward \\ \midrule
            100         & 2.0 & 2.0     & 2.0     \\
            1000        & 3.3 & 3.5     & 3.7     \\
            2500        & 4.2 & 4.3     & 4.7     \\
            5000        & 4.5 & --      & --      \\ \bottomrule
        \end{tabular}
        \caption{Varying endocardial pressure.}
    \end{subtable}
    \begin{subtable}{0.46\textwidth}
        \centering
        \begin{tabular}{|r|r|r|r|r|}
            \toprule Pa & ID  & ID-forw & Forward \\ \midrule
            1000        & 3.0 & 3.0     & 3.0     \\
            5000        & 3.2 & 3.0     & 3.7     \\
            10000       & 3.4 & 3.4     & 3.8     \\
            20000       & 3.8 & 3.6     & 3.9     \\ \bottomrule
        \end{tabular}
        \caption{Varying active stress.}
    \end{subtable}

    \caption{Computational effort study on LV. (ID) Inverse displacement problem, (ID-forw) Forward problem from stress-free configuration, and (Forward) is the forward problem from the current configuration.}
    \label{table:comparison-lv}
\end{table}

\subsection{Preconditioning}\label{section:preconditioning}
The main strategy so far to compute preconditioners for nonlinear elasticity has been to devise optimal preconditioners for the linearized formulation, and then use such techniques for the nonlinear scenario. This usually yields satisfactory results in the nonlinear regime, but in this section we show that this approach is not equally valid for the inverse displacement problem. For this, consider the slab problem shown in Section~\ref{sec:beam-tests} with a volumetric load given by $\vec b=-\SI{32}{\pascal\per\meter}\vec{e}_3$, which we have observed to be sufficiently large to challenge the numerical solvers. In contrast to all previous numerical tests, where we have used a direct solver (MUMPS) for all linear systems, we display the average number of GMRES iterations using the well-established Algebraic Multigrid implementation from HYPRE~\cite{falgout2002hypre} for an increasing number of degrees of freedom. This solver is an excellent choice for nonlinear elasticity, and its efficiency has been thoroughly studied for cardiac elasticity as well~\cite{barnafi2022comparative}.

\begin{figure}[ht!]
    \centering
    
    \begin{tikzpicture}
    \begin{loglogaxis}
        [width=0.7\textwidth, height=6cm, xlabel=DoFs, ylabel=, tick label style={font=\normalsize},line width=2pt,legend style={draw=none, fill opacity=0.7, text opacity= 1,row sep =2pt, font=\normalsize, at={(0.95,0.7)}, anchor=east}, legend columns=1, cycle list/Set1-5]
        \addplot+[mark=square*] coordinates {
                                                    (975,30)
                                                    (6075, 55)
                                                    (42483, 101)
                                                    };
        \addplot+[mark=square*] coordinates {
                                                    (975,62) 
                                                    (6075, 118)
                                                    (42483, 287)
                                                    (316889, 1418)
                                                    };
        \addplot+[mark=square*] coordinates {
                                                    (975,14) 
                                                    (6075, 18)
                                                    (42483, 29)
                                                    (316889, 53)
                                                    (2446275, 100)
                                                    };
        \legend{AMG, AS, GDSW}
    \end{loglogaxis}
    \end{tikzpicture}

    \caption{Average number of GMRES iterations incurred in one ramp step by each of the preconditioners considered: (AMG) Algebraic Multigrid, (AS) Additive Schwarz with ILU on 8 sub-domains, and (GDSW) Generalized Driya-Smith-Widlund. Unplotted points mean that the solver attained 5000 linear iterations.} %
    \label{table:preconditioning}
\end{figure}
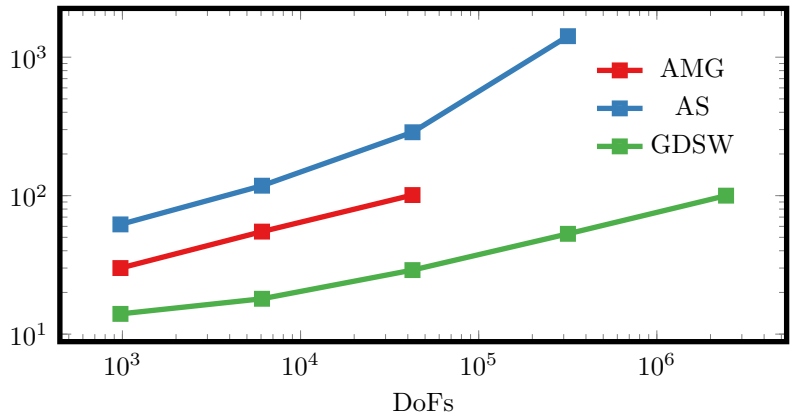

We compare its performance with a very simple one-level Additive Schwarz preconditioner with minimal overlap and an incomplete LU (ILU) factorization as a local solver, and show the performance with 16 subdomains (16 MPI processes) in Figure~\ref{table:preconditioning}. We see that, surprisingly, AMG is particularly not suitable for this problem, and an AS/ILU preconditioner provides a better alternative. Moreover, we test the GDSW preconditioner~\cite{dohrmann2008family} available in PETSc~\cite{petsc-user-ref} under the same conditions. We provide the results in Figure~\ref{table:preconditioning}, where a much improved performance is obtained in terms of linear iterations, albeit not optimal. We report the PETSc options to use these preconditioners in~\ref{appendix:petsc}. We note that none of the tested preconditioners are optimal, and that obtaining an optimal preconditioner, at least in practice, for the inverse displacement formulation is beyond the scope of this work.

\subsection{Realistic four-chamber heart} \label{sec:cardio_results}

\begin{figure}[ht!]
    \centering
    \includegraphics[width=1\textwidth]{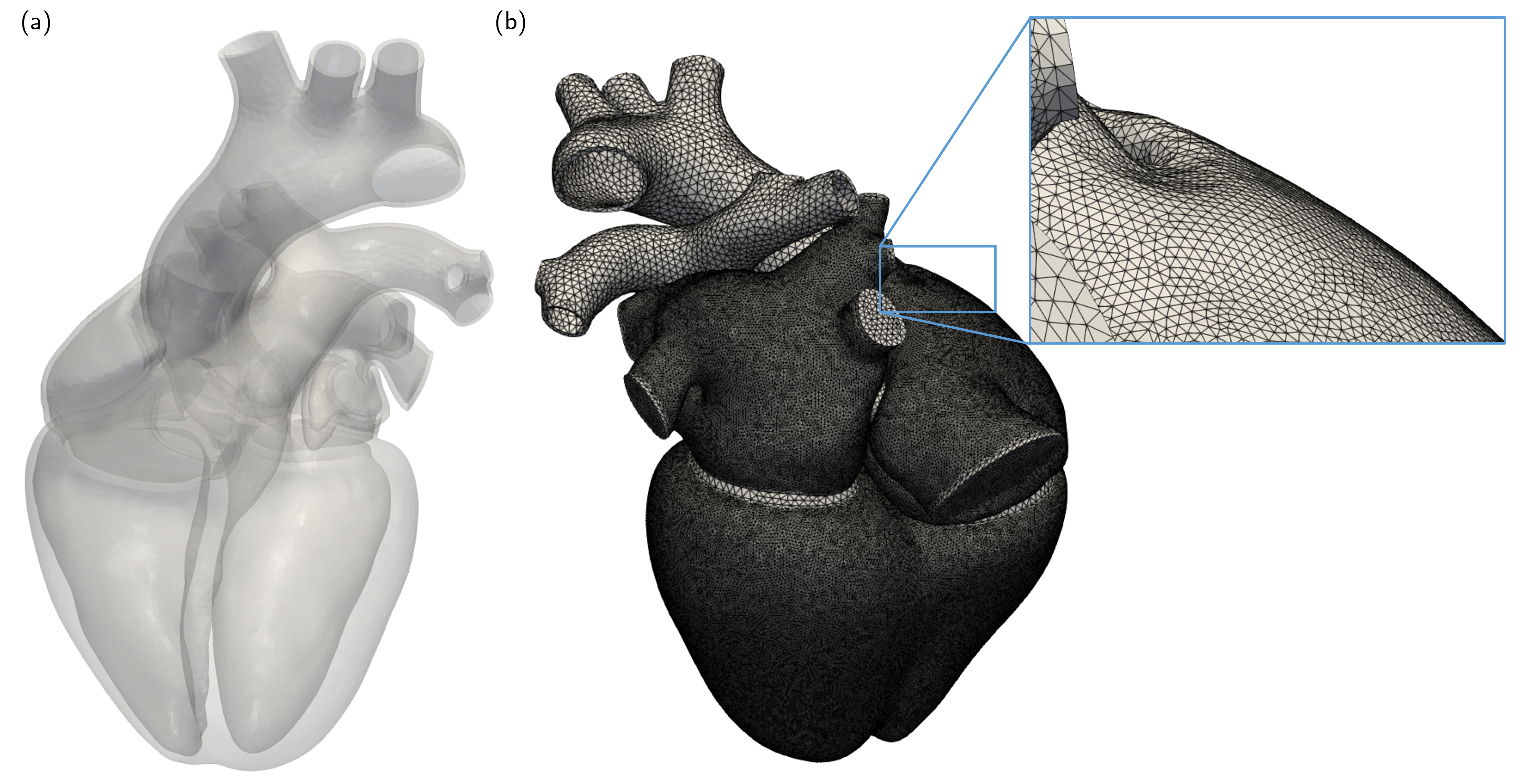}
    \caption{Full heart model: (a) computational domain and (b) computational mesh. The mesh is finer in the conductive regions of the myocardium.}
    \label{fig:cardio_domain_mesh}
\end{figure}

We now turn to the case of a realistic full-heart model, introduced in Section~\ref{sec:cardio_model}.
We consider the Zygote Solid 3D Heart Model~\cite{Zygote2014}, an anatomically accurate CAD model of the whole human heart, obtained from high-resolution CT scans and representing an average healthy male subject, displayed in Figure~\ref{fig:cardio_domain_mesh}a.
We consider a computational mesh with an average cell diameter of $\SI{1.18}{\milli\meter}$ and accounting for \num{2.75e6} tetrahedra (see Figure~\ref{fig:cardio_domain_mesh}b), generated relying on the algorithms proposed in Ref.~\cite{Fedele2021}, implemented in the open source software \texttt{vmtk}~\cite{Antiga2008}.

We generate the fiber architecture by relying on the Laplace-Dirichlet rule-based method for whole heart geometries proposed in Ref.~\cite{piersanti2021modeling} and further refined in Ref.~\cite{fedele2023comprehensive}.
In the myocardium, we consider the exponential constitutive law of Usyk~\cite{Usyk2002}, with a volumetric term enforcing quasi-incompressibility:
\begin{equation*}
    \Psi(\ten F ) = \dfrac{C}{2} \left( e^Q  - 1 \right) + \dfrac{B}{2} \left( J - 1 \right) \log(J)
\end{equation*}
where $C$ is the material stiffness, $B$ is the bulk modulus, and
\begin{equation*}
    \begin{aligned}
         & Q =
        b_{\mathrm{ff}} E_{\mathrm{ff}}^2  + b_{\mathrm{ss}} E_{\mathrm{ss}}^2 + b_{\mathrm{nn}} E_{\mathrm{nn}}^2+ b_{\mathrm{fs}} \left( E_{\mathrm{fs}}^2 + E_{\mathrm{sf}}^2 \right) + b_{\mathrm{fn}} \left( E_{\mathrm{fn}}^2 + E_{\mathrm{nf}}^2 \right) + b_{\mathrm{sn}} \left( E_{\mathrm{sn}}^2 + E_{\mathrm{ns}}^2 \right), \\
         & E_\text{ab} =
        \ten E \fiberRef{a} \cdot \fiberRef{b}, \qquad \text{for }a, b \in \{ f, s, n \},
    \end{aligned}
\end{equation*}
where $\ten E = \tfrac{1}{2} \left( \ten C - \ten I \right)$ is the Green-Lagrange strain energy tensor, being $\ten C = \ten F^{T} \ten F$ the right Cauchy-Green deformation tensor. See the aforementioned reference for the parameter values. In the vessels, instead, we use the Neo-Hookean model:
\begin{equation*}
    \Psi(\ten F ) = \frac{\mu}{2}\left(J^{-\frac{2}{3}}\tr (\ten F^T\ten F) -3\right) + \dfrac{\kappa}{4}\left[ \left( J - 1 \right)^2 + \log^2(J)\right].
\end{equation*}
We employ the parameters values reported in Ref.~\cite{fedele2023comprehensive}.
We consider a factor $\ratioN = 0.4$ to account for the effect of microscale fiber dispersion on the active stress.
Concerning the epicardium boundary conditions, we set $\BCKt = 0$ and $\BCKn = \SI{2e5}{\pascal\per\meter}$.

To define the IEP, we consider the pressures and diastolic tension reported in Table~\ref{table:cardio-load}.
We consider two load cases, namely 50\% and 100\% of the values reported in Table~\ref{table:cardio-load}.

\begin{table}
    \centering
    \begin{tabular}{lr}
        \toprule
        Left ventricle   pressure & \SI{13.1}{\mmHg}       \\
        Right ventricle  pressure & \SI{ 6.0}{\mmHg}       \\
        Left atrium      pressure & \SI{ 9.0}{\mmHg}       \\
        Right atrium     pressure & \SI{ 5.6}{\mmHg}       \\
        Ascending aorta  pressure & \SI{71.2}{\mmHg}       \\
        Pulmonary trunk  pressure & \SI{12.8}{\mmHg}       \\
        Active tension            & \SI{7.5}{\kilo\pascal} \\
        \bottomrule
    \end{tabular}
    \caption{Load values considered in the full-heart test case.}
    \label{table:cardio-load}
\end{table}

In this test case, we consider both the IEP (that we address with the inverse displacement method and with the Sellier method), and the DEP after having computed the stress-free configuration.
Both problems are considerably challenging, due to the highly-nonlinear constitutive law and to the nontrivial geometric features of the considered domain.
As a matter of fact, none of the solution methods considered is able to reach convergence with a single load step.
Hence, we consider a load-ramp approach by increasing simultaneously both the cavity pressures and the active tension.
In such a challenging problem, as the ramp approaches the target value, smaller and smaller steps are typically required to avoid convergence failures, especially when the Sellier method is considered~\cite{regazzoni2022emcirculation}.
Hence, to avoid the need of manually tuning the ramp step, we implement the adaptive ramp algorithm of~\cite{regazzoni2022emcirculation}, by which the ramp step size is automatically decreased (by a factor 0.7) in case of failure, while it is increased (by a factor 1.2, with a maximum of 0.2 relative step length) in case of success.

The considered problem is challenging also because of the ill-conditioning of the linear systems arising from each Newton iteration.
Hence, in order to mitigate the computational burden, we consider, besides the standard Newton algorithm, an Inexact Newton algorithm that employs a loose tolerance for the linear solver in the first nonlinear solver iteration, and progressively reduces it during the iterations. This strategy has been shown to reduce CPU times in cardiac simulations~\cite{barnafi2022parallel} without sacrificing robustness nor optimality.
Moreover, we employ the GMRES method by setting a large maximum number of iterations (namely $10^{4}$).
We consider both an algebraic multigrid (AMG) preconditioner, and an additive Schwarz method with an ILU approximate solve as inner solver, based on the parallel partitioning (AS/ILU) as shown in Section~\ref{section:preconditioning}. For linear algebra operations we rely on the Trilinos library~\cite{trilinos-website}. Simulations are run on a parallel computing cluster on 92 cores (Lenovo SR950 192-Core Intel Xeon Platinum 8160, 2100 MHz and 1.7TB RAM) at MOX, Department of Mathematics, Politecnico di Milano.

We report in Table~\ref{table:cardio-results} the results in terms of convergence success, wall time, number of iterations.
First, we notice that the Inexact Newton approach brings in all the considered cases a significant advantage (between 2x and 8x speedup).

\begin{sidewaystable}
    \centering
    \begin{tabular}{|lll|r>{\raggedleft\arraybackslash}p{6mm}>{\raggedleft\arraybackslash}p{6mm}>{\raggedleft\arraybackslash}p{12mm}|r>{\raggedleft\arraybackslash}p{6mm}>{\raggedleft\arraybackslash}p{6mm}>{\raggedleft\arraybackslash}p{12mm}|}
        \toprule
                &                 &              & \multicolumn{4}{c|}{\textbf{50\% load}} & \multicolumn{4}{c|}{\textbf{100\% load}}                                                                                      \\
        Problem & Nonlinear solv. & Linear solv. & Wall time                               & F.P. steps                               & Newt. steps & Time per step & Wall time & F.P. steps & Newt. steps & Time per step \\
        \midrule
        ID      & Newton          & AMG          & $>$ 24h                                 &                                          &             &               & $>$ 24h   &            &             &               \\
        ID      & Inexact Newton  & AMG          & $>$ 24h                                 &                                          &             &               & $>$ 24h   &            &             &               \\
        ID      & Newton          & AS/ILU       & 36m 20s                                 &                                          & 21          & 103.8s        & 64m 20s   &            & 43          & 89.8s         \\
        ID      & Inexact Newton  & AS/ILU       & 12m 01s                                 &                                          & 25          & 28.8s         & 19m 30s   &            & 38          & 30.8s         \\ \midrule
        Sellier & Newton          & AMG          & 331m 41s                                & 82                                       & 295         & 67.5s         & $>$ 24h   &            &             &               \\
        Sellier & Inexact Newton  & AMG          & 41m 00s                                 & 42                                       & 160         & 15.4s         & $>$ 24h   &            &             &               \\
        Sellier & Newton          & AS/ILU       & 568m 20s                                & 108                                      & 364         & 93.7s         & $>$ 24h   &            &             &               \\
        Sellier & Inexact Newton  & AS/ILU       & 77m 32s                                 & 42                                       & 166         & 28.1s         & $>$ 24h   &            &             &               \\ \midrule
        ID-forw & Newton          & AMG          & 20m 31s                                 &                                          & 27          & 45.6s         & 32m 02s   &            & 40          & 48.0s         \\
        ID-forw & Inexact Newton  & AMG          & 7m 10s                                  &                                          & 25          & 17.2s         & 13m 51s   &            & 41          & 20.3s         \\
        ID-forw & Newton          & AS/ILU       & 42m 50s                                 &                                          & 27          & 95.2s         & 40m 50s   &            & 40          & 61.3s         \\
        ID-forw & Inexact Newton  & AS/ILU       & 14m 06s                                 &                                          & 26          & 32.5s         & 20m 04s   &            & 42          & 28.6s         \\
        \bottomrule
    \end{tabular}
    \caption{Results of the realistic 4 chamber cardiac model of Section~\ref{sec:cardio_results}. We report:
        the wall time;
        the number of fixed point steps (only for the Sellier method);
        the total number of Newton steps (summed over the ramp steps and, for the Sellier method, over the fixed point iterations);
        The wall time per each Newton step.
    }
    \label{table:cardio-results}
\end{sidewaystable}

Instead, the two considered preconditioners behave very differently depending on the differential problem being solved (namely \eqref{eq:forward weak} or \eqref{eq:backward weak}).
As shown in Section~\ref{section:preconditioning}, AMG shows to be ineffective for the IEP, since the maximum number of GMRES iterations is very often reached, despite the fact that the adaptive algorithm leads to smaller and smaller steps in the ramps.  Instead, the AS/ILU method provides a more robust preconditioner for this problem.
In contrast, when we consider the DEP or the Sellier method for the IEP (which, at each step, solves a DEP), the choice of preconditioner does not determine the ability to reach convergence or not, but it impacts the wall time.
Comparing the results obtained with AMG and AS/ILU, we see that the number of Newton steps is virtually identical, but the wall time is roughly half using AMG.
The only exception is in the case of the Sellier method with the traditional Newton algorithm, for which using AS/ILU the nonlinear solver performs about 50\% more iterations, meaning that GMRES fails more often than with AMG.
In any case, for solving the DEP \eqref{eq:forward weak}, AMG proves preferable to AS/ILU.

Finally, we compare the inverse displacement method with the Sellier method in solving the IEP.
We observe that the Sellier method is unable to reach convergence in this real-life test case when 100\% of the load is considered, regardless the nonlinear and linear solvers employed.
When we consider a 50\% reduction of the load, both methods converge, but the inverse displacement method is remarkably more efficient (12 minutes against 41 minutes in the best case, that is with AS/ILU and AMG, respectively).
The inverse displacement requires the resolution of more demanding linear systems (28.8s against 15.4s), but this is compensated by a significantly smaller number of Newton steps (25 against 160).

\begin{figure}[ht!]
    \centering
    \includegraphics[width=1\textwidth]{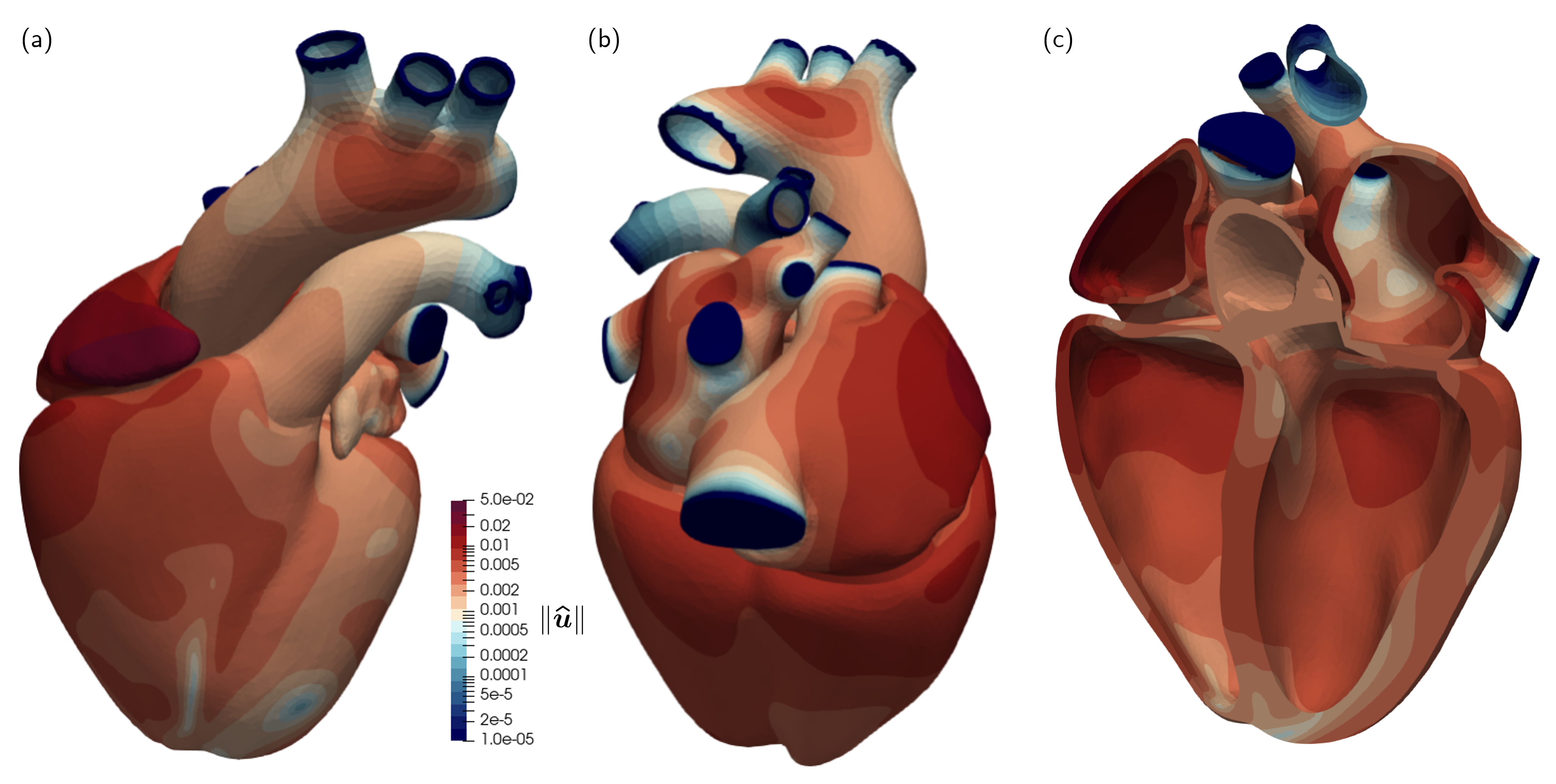}
    \caption{Inverse displacement magnitude in the full heart test case.
        We report:
        (a) a frontal view;
        (b) a top view showing the atria;
        (c) a section, showing the endocardium.}
    \label{fig:cardio_magnitude}
\end{figure}

\begin{figure}[ht!]
    \centering
    \includegraphics[width=1\textwidth]{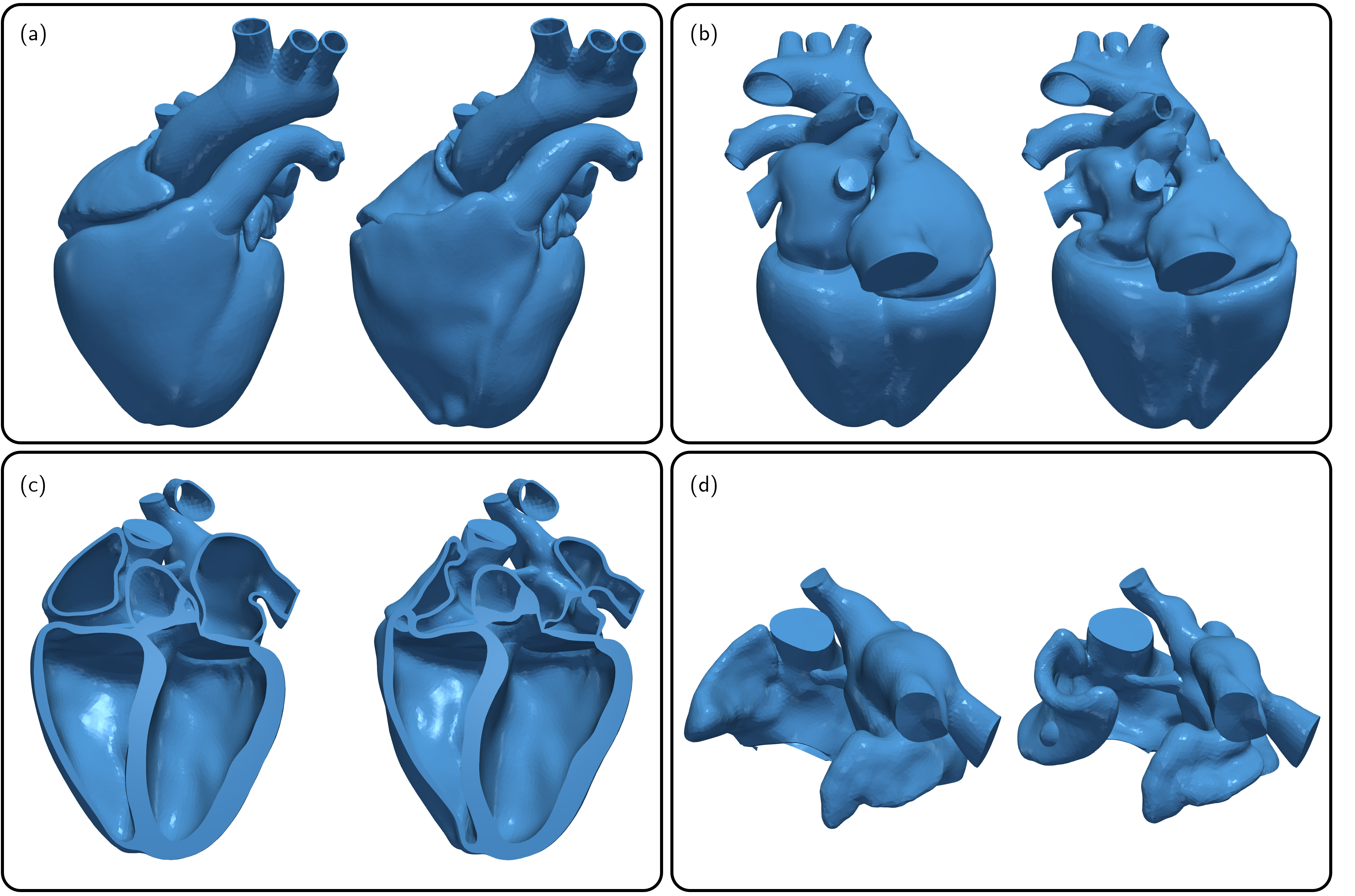}
    \caption{Original domain (left) and stress-free configuration (right) in the full heart test case.
        We show:
        (a) a frontal view;
        (b) a top view;
        (c) a section, showing the endocardium;
        (d) a section, showing the atria.}
    \label{fig:cardio_comparison}
\end{figure}

We conclude this section by showing the results obtained in the real-life full heart model.
In Figure~\ref{fig:cardio_magnitude} we show the magnitude of the displacement from the stress-free configuration and the deformed one.
In Figure~\ref{fig:cardio_comparison} we report several views of the deformed and the stress-free configuration.
As expected, the cardiac chambers are deflated, because of the pressures acting on the endocardium.
In addition, the chambers that are deformed the most are those with a thinner wall, since they are more prone to being stretched by pressure, and thus the stress-free configuration is more distant from the deformed one.
Atria are deflated to a remarkable degree, an aspect that makes calculating the stress-free configuration particularly challenging in this test case.
Such deflation induces a rotation in the right atrium auricle, so that a self-penetration of the domain occurs, both of the atrium into the ventricle and of the opposite walls of the atrium.
The self-penetration of the relaxed configuration is the manifestation of a global geometric incompatibility, as discussed in Section~\ref{sec:self-penetration}.

\section{Conclusions}\label{section:conclusions}

In this paper, we have delved into the complex task of reconstructing the stress-free configuration of an elastic body, terming this challenge the \emph{inverse elasticity problem}.

In Section~\ref{section:problem-description}, we have demonstrated that obtaining the inverse deformation map involves solving a mixed boundary value problem that shares structural similarities with the classical problem of hyperelasticity. Expanding upon Shield's pioneering findings~\cite{Shield_1967}, we have extended our analysis to encompass the impact of material inhomogeneities, body and active forces. In particular, a modification of the balance equation is necessary to account for material inhomogeneities. Both body forces and material inhomogeneities can break the variational structure of the problem.

In this respect, our investigation has revealed that the existence of solutions can be ensured under stringent assumptions, we have uncovered that, even for a simple scenario involving a two-dimensional disk composed of Neo-Hookean material and subjected to external pressure, the problem can yield one, multiple, or even zero solutions depending on the applied pressure.

Furthermore, we have conducted an analysis of potential global geometric incompatibilities, leading to a non-injective inverse deformation. While injectivity of the deformation is pivotal in the direct problem to avoid self-intersections, we have shown that this characteristic is not mandatory for the inverse deformation, and characterized numerically two different mechanisms in which this phenomenon can arise. Nevertheless, the resulting self-intersecting relaxed state of the body could pose issues, rendering the domain unsuitable as a reference configuration. To counteract this challenge, we have proposed a novel approach outlined in Section~\ref{sec:self-penetration}, based on a multiplicative decomposition of the deformation gradient tensor.

We have then thoroughly studied the inverse displacement method in terms of its numerical behavior, both independently and in comparison to alternative fixed-point (Sellier) algorithms. Our numerical evidence suggests that
\begin{itemize}
    \item[(i)] the inverse displacement method outperforms the Sellier methods in terms of convergence speed and robustness,
    \item[(ii)] the Sellier algorithm can be slightly enhanced with Anderson acceleration, but the advantage is negligible when compared against using the inverse displacement method,
    \item[(iii)] the inverse displacement problem can be equivalently formulated in terms of the Cauchy and Eshelby stress tensors, but using the Cauchy formulation requires a smaller computational effort,
    \item[(iv)] in terms of nonlinear solvers, the inverse displacement problem behaves similarly to the standard elasticity problem, and
    \item[(v)] preconditioning the inverse displacement problem is significantly more challenging, and we have shown that domain decomposition preconditioners are significantly more effective than AMG.
\end{itemize}

We have challenged both the inverse displacement method and the Sellier method in a real-life full heart test case, characterized by detailed anatomical features and by a computational mesh having \num{2.75e6} tetrahedra.
The most noticeable result of this test is the greater robustness and better performance of the inverse displacement method compared with Sellier's method, which is the one currently most widely used -- to our knowledge -- in the cardiac modeling community.
As a matter of fact, by relying on the inverse displacement method we were able to recover the stress-free configuration for a realistic load. %
Remarkably, the computation took only 19m 30s on 92 cores, that is only 40\% more than solving the direct elasticity problem on the same mesh for the same load.

\appendix
\section{PETSc options for preconditioners}\label{appendix:petsc}
In Listings~\ref{lis:AMG}-\ref{lis:GDSW}, we report the PETSc options used to test the preconditioners AMG, ILU, and GDSW in Section~\ref{section:preconditioning}.

\begin{lstlisting}[language=python, float, frame=single, caption=PETSc commands to use AMG., label=lis:AMG]    
    "snes_type": "newtonls",
    "snes_atol": 1e-12,
    "snes_rtol": 1e-6,
    "snes_stol": 0.0,
    "snes_linesearch_type": "basic",
    "ksp_type": "gmres",
    "ksp_atol": 0.0,
    "ksp_rtol": 1e-6,
    "ksp_max_it": 5000,
    "ksp_norm_type": "unpreconditioned",
    "ksp_gmres_restart": 1000,
    "pc_type": "hypre"
  \end{lstlisting}

\begin{lstlisting}[language=python, frame=single, float,caption=PETSc commands to use AS., label = lis:ILU]     
    "snes_type": "newtonls",
    "snes_atol": 1e-12,
    "snes_rtol": 1e-6,
    "snes_stol": 0.0,
    "snes_linesearch_type": "basic",
    "ksp_type": "gmres",
    "ksp_atol": 0.0,
    "ksp_rtol": 1e-6,
    "ksp_max_it": 5000,
    "ksp_norm_type": "unpreconditioned",
    "ksp_gmres_restart": 1000,
    "pc_type": "asm",
    "sub_ksp_type": "preonly",
    "sub_pc_type": "ilu"
  \end{lstlisting}

\begin{lstlisting}[language=python, frame=single, float, label = lis:GDSW, caption=PETSc commands to use GDSW.] 
    "snes_type": "newtonls",
    "snes_atol": 1e-12,
    "snes_rtol": 1e-6,
    "snes_stol": 0.0,
    "snes_linesearch_type": "basic",
    "ksp_type": "gmres",
    "ksp_atol": 0.0,
    "ksp_rtol": 1e-6,
    "ksp_max_it": 5000,
    "ksp_norm_type": "unpreconditioned",
    "ksp_gmres_restart": 1000,
    "pc_type": "mg",
    "pc_mg_galerkin": None,
    "pc_mg_levels": 2,
    "pc_mg_adapt_interp_coarse_space": "gdsw",
    "mg_levels_pc_type": "asm"
  \end{lstlisting}

\section*{Acknowledgment}
DR gratefully acknowledges funding by the European Union – NextGenerationEU under the National Recovery and Resilience Plan (NRRP), Mission 4 Component 2 Investment 1.1 - Call PRIN 2022 No. 104 of February 2, 2022 of Italian Ministry of University and Research; Project 202249PF73 (subject area: PE - Physical Sciences and Engineering) ``Mathematical models for viscoelastic biological matter''. FR has received support from the project PRIN2022, MUR, Italy, 2023--2025, P2022N5ZNP ``SIDDMs: shape-informed data-driven models for parametrized PDEs, with application to computational cardiology''. NB has been supported by CMM BASAL proyect FB2100005 and by ANID POSTDOCTORAL 3230325. FR and DR are members of the INdAM research group GNCS (FR) and GNFM (DR).
FR and DR acknowledge the support by the MUR, Italian Ministry of University and Research (Italy), grant ``Dipartimento di Eccellenza 2023-2027''.


\begin{thebibliography}{10}

\bibitem{africa2022lifex}
P.~C. Africa.
\newblock lifex: A flexible, high performance library for the numerical
  solution of complex finite element problems.
\newblock {\em SoftwareX}, 20:101252, 2022.

\bibitem{alnaes2015fenics}
M.~Aln{\ae}s, J.~Blechta, J.~Hake, A.~Johansson, B.~Kehlet, A.~Logg,
  C.~Richardson, J.~Ring, M.~Rognes, and G.~Wells.
\newblock The {FEniCS} project version 1.5.
\newblock {\em Archive of Numerical Software}, 3(100), 2015.

\bibitem{Aln_s_2014}
M.~S. Alnæs, A.~Logg, K.~B. Ølgaard, M.~E. Rognes, and G.~N. Wells.
\newblock Unified form language: A domain-specific language for weak
  formulations of partial differential equations.
\newblock {\em ACM Transactions on Mathematical Software}, 40(2):1–37, Feb.
  2014.

\bibitem{Ambrosi_2011}
D.~Ambrosi and S.~Pezzuto.
\newblock Active stress vs. active strain in mechanobiology: Constitutive
  issues.
\newblock {\em Journal of Elasticity}, 107(2):199–212, July 2011.

\bibitem{amestoy2000mumps}
P.~R. Amestoy, I.~S. Duff, J.-Y. L’Excellent, and J.~Koster.
\newblock {MUMPS}: a general purpose distributed memory sparse solver.
\newblock In {\em International Workshop on Applied Parallel Computing}, pages
  121--130. Springer, 2000.

\bibitem{Antiga2008}
L.~Antiga, M.~Piccinelli, L.~Botti, B.~Ene-Iordache, A.~Remuzzi, and D.~A.
  Steinman.
\newblock An image-based modeling framework for patient-specific computational
  hemodynamics.
\newblock {\em Medical \& Biological Engineering \& Computing}, 46(11), nov
  2008.

\bibitem{dealII91}
D.~Arndt, W.~Bangerth, T.~Clevenger, D.~Davydov, M.~Fehling, D.~Garcia-Sanchez,
  G.~Harper, T.~Heister, L.~Heltai, M.~Kronbichler, R.~Kynch, M.~Maier, J.-P.
  Pelteret, B.~Turcksin, and D.~Wells.
\newblock {The \texttt{deal.II} Library, Version 9.1}.
\newblock {\em Journal of Numerical Mathematics}, 2019.

\bibitem{petsc-user-ref}
S.~Balay, S.~Abhyankar, M.~Adams, J.~Brown, P.~Brune, K.~Buschelman, L.~Dalcin,
  A.~Dener, V.~Eijkhout, W.~Gropp, D.~Karpeyev, D.~Kaushik, M.~Knepley, D.~May,
  L.~Curfman~McInnes, R.~Mills, T.~Munson, K.~Rupp, P.~Sanan, B.~Smith,
  S.~Zampini, H.~Zhang, and H.~Zhang.
\newblock {PETS}c users manual.
\newblock Technical Report ANL-95/11 - Revision 3.13, Argonne National
  Laboratory, 2021.

\bibitem{ball1976convexity}
J.~Ball.
\newblock Convexity conditions and existence theorems in nonlinear elasticity.
\newblock {\em Archive for Rational Mechanics and Analysis}, 63(4):337--403,
  1976.

\bibitem{barnafi2022parallel}
N.~A. Barnafi, L.~F. Pavarino, and S.~Scacchi.
\newblock Parallel inexact newton--krylov and quasi-newton solvers for
  nonlinear elasticity.
\newblock {\em Computer Methods in Applied Mechanics and Engineering},
  400:115557, 2022.

\bibitem{barnafi2022comparative}
N.~A. Barnafi, L.~F. Pavarino, and S.~Scacchi.
\newblock A comparative study of scalable multilevel preconditioners for
  cardiac mechanics.
\newblock {\em Journal of Computational Physics}, 492:112421, 2023.

\bibitem{bayer2012novel}
J.~D. Bayer, R.~C. Blake, G.~Plank, and N.~A. Trayanova.
\newblock A novel rule-based algorithm for assigning myocardial fiber
  orientation to computational heart models.
\newblock {\em Annals of Biomedical Engineering}, 40:2243--2254, 2012.

\bibitem{bucelli2022partitioned}
M.~Bucelli, L.~Dede, A.~Quarteroni, and C.~Vergara.
\newblock Partitioned and monolithic algorithms for the numerical solution of
  cardiac fluid-structure interaction.
\newblock {\em Communications in Computational Physics}, 32(5):1217--1256, jun
  2022.

\bibitem{carroll2005compressible}
M.~M. Carroll.
\newblock Compressible isotropic strain energies that support universal
  irrotational finite deformations.
\newblock {\em The Quarterly Journal of Mechanics and Applied Mathematics},
  58(4):601--614, 2005.

\bibitem{Carroll_2005}
M.~M. Carroll and F.~J. Rooney.
\newblock Implications of shield's inverse deformation theorem for compressible
  finite elasticity.
\newblock {\em Zeitschrift für angewandte Mathematik und Physik},
  56(6):1048--1060, nov 2005.

\bibitem{Chadwick_1975}
P.~Chadwick.
\newblock Applications of an energy-momentum tensor in non-linear
  elastostatics.
\newblock {\em Journal of Elasticity}, 5(3-4):249--258, nov 1975.

\bibitem{deng2023fast}
G.~Deng and F.~Galetto.
\newblock Fast iterative reverse filters using fixed-point acceleration.
\newblock {\em Signal, Image and Video Processing}, pages 1--9, 2023.

\bibitem{dicarlo2002growth}
A.~DiCarlo and S.~Quiligotti.
\newblock Growth and balance.
\newblock {\em Mechanics Research Communications}, 29(6):449--456, 2002.

\bibitem{dohrmann2008family}
C.~R. Dohrmann, A.~Klawonn, and O.~B. Widlund.
\newblock A family of energy minimizing coarse spaces for overlapping schwarz
  preconditioners.
\newblock In {\em Domain Decomposition Methods in Science and Engineering
  XVII}, pages 247--254. Springer, 2008.

\bibitem{epstein2015mathematical}
M.~Epstein.
\newblock Mathematical characterization and identification of remodeling,
  growth, aging and morphogenesis.
\newblock {\em Journal of the Mechanics and Physics of Solids}, 84:72--84,
  2015.

\bibitem{evans2020proof}
C.~Evans, S.~Pollock, L.~G. Rebholz, and M.~Xiao.
\newblock A proof that anderson acceleration improves the convergence rate in
  linearly converging fixed-point methods (but not in those converging
  quadratically).
\newblock {\em SIAM Journal on Numerical Analysis}, 58(1):788--810, 2020.

\bibitem{falgout2002hypre}
R.~Falgout and U.~Yang.
\newblock hypre: A library of high performance preconditioners.
\newblock In {\em International Conference on Computational Science}, pages
  632--641. Springer, 2002.

\bibitem{fedele2023comprehensive}
M.~Fedele, R.~Piersanti, F.~Regazzoni, M.~Salvador, P.~C. Africa, M.~Bucelli,
  A.~Zingaro, A.~Quarteroni, et~al.
\newblock A comprehensive and biophysically detailed computational model of the
  whole human heart electromechanics.
\newblock {\em Computer Methods in Applied Mechanics and Engineering},
  410:115983, 2023.

\bibitem{Fedele2021}
M.~Fedele and A.~Quarteroni.
\newblock Polygonal surface processing and mesh generation tools for the
  numerical simulation of the cardiac function.
\newblock {\em International Journal for Numerical Methods in Biomedical
  Engineering}, 37(4):e3435, 2021.

\bibitem{gee2009prestressing}
M.~W. Gee, C.~H. Reeps, H.~H. Eckstein, and W.~A. Wall.
\newblock Prestressing in finite deformation abdominal aortic aneurysm
  simulation.
\newblock {\em Journal of Biomechanics}, 42(11):1732--1739, 2009.

\bibitem{genet_2023_10299077}
M.~Genet.
\newblock dolfin\_mech, Dec. 2023.

\bibitem{genet2015heterogeneous}
M.~Genet, M.~Rausch, L.~C. Lee, S.~Choy, X.~Zhao, G.~S. Kassab, S.~Kozerke,
  J.~M. Guccione, and E.~Kuhl.
\newblock Heterogeneous growth-induced prestrain in the heart.
\newblock {\em Journal of biomechanics}, 48(10):2080--2089, 2015.

\bibitem{Giantesio_2018}
G.~Giantesio, A.~Musesti, and D.~Riccobelli.
\newblock A comparison between active strain and active stress in transversely
  isotropic hyperelastic materials.
\newblock {\em Journal of Elasticity}, 137(1):63--82, dec 2018.

\bibitem{Goriely_2005}
A.~Goriely and M.~Ben~Amar.
\newblock Differential growth and instability in elastic shells.
\newblock {\em Physical Review Letters}, 94(19), May 2005.

\bibitem{govindjee1996computational}
S.~Govindjee and P.~A. Mihalic.
\newblock Computational methods for inverse finite elastostatics.
\newblock {\em Computer Methods in Applied Mechanics and Engineering},
  136(1-2):47--57, 1996.

\bibitem{henderson2007view}
A.~Henderson.
\newblock Paraview guide: A parallel visualization application.
\newblock {\em Kitware, Inc., Clifton Park, NY}, 2007.

\bibitem{horgan2004invariance}
C.~O. Horgan and J.~G. Murphy.
\newblock Invariance of the equilibrium equations of finite elasticity for
  compressible materials.
\newblock {\em Journal of Elasticity}, 77:187--200, 2004.

\bibitem{horgan2005plane}
C.~O. Horgan and J.~G. Murphy.
\newblock Plane strain bending of cylindrical sectors of admissible
  compressible hyperelastic materials.
\newblock {\em Journal of Elasticity}, 81:129--151, 2005.

\bibitem{katz2010}
A.~M. Katz.
\newblock {\em Physiology of the Heart}.
\newblock Lippincott Williams \& Wilkins, 2010.

\bibitem{kelley1991mesh}
C.~Kelley and E.~W. Sachs.
\newblock Mesh independence of newton-like methods for infinite dimensional
  problems.
\newblock {\em The Journal of Integral Equations and Applications}, pages
  549--573, 1991.

\bibitem{kondaurov1987finite}
V.~I. Kondaurov and L.~V. Nikitin.
\newblock Finite strains of viscoelastic muscle tissue.
\newblock {\em Journal of Applied Mathematics and Mechanics}, 51(3):346--353,
  1987.

\bibitem{kroner1959allgemeine}
E.~Kr{\"o}ner.
\newblock Allgemeine kontinuumstheorie der versetzungen und eigenspannungen.
\newblock {\em Archive for Rational Mechanics and Analysis}, 4(1):273--334,
  1959.

\bibitem{Lee1969}
E.~H. Lee.
\newblock Elastic-plastic deformation at finite strains.
\newblock {\em Journal of Applied Mechanics}, 36(1):1--6, Mar 1969.

\bibitem{marx2022robust}
L.~Marx, J.~A. Niestrawska, M.~A.~F. Gsell, F.~Caforio, G.~Plank, and C.~M.
  Augustin.
\newblock Robust and efficient fixed-point algorithm for the inverse
  elastostatic problem to identify myocardial passive material parameters and
  the unloaded reference configuration.
\newblock {\em Journal of Computational Physics}, 463:111266, 2022.

\bibitem{mazier2022inverse}
A.~Mazier, A.~Bilger, A.~E. Forte, I.~Peterlik, J.~S. Hale, and S.~P. Bordas.
\newblock Inverse deformation analysis: an experimental and numerical
  assessment using the {FEniCS Project}.
\newblock {\em Engineering with Computers}, 38(5):4099--4113, 2022.

\bibitem{Merodio_2006}
J.~Merodio and R.~W. Ogden.
\newblock On the equivalence of strong ellipticity in the material and spatial
  settings of finite elasticity.
\newblock {\em Zeitschrift für Angewandte Mathematik und Physik},
  57(6):1096--1101, aug 2006.

\bibitem{montanino2020recovery}
A.~Montanino and A.~Pandolfi.
\newblock On the recovery of the stress-free configuration of the human cornea.
\newblock {\em Journal for Modeling in Ophthalmology}, 4:11--33, 2020.

\bibitem{mora2019shape}
S.~Mora, E.~And{\`o}, J.-M. Fromental, T.~Phou, and Y.~Pomeau.
\newblock The shape of hanging elastic cylinders.
\newblock {\em Soft Matter}, 15(27):5464--5473, 2019.

\bibitem{Mora_2014}
S.~Mora, T.~Phou, J.-M. Fromental, and Y.~Pomeau.
\newblock Gravity driven instability in elastic solid layers.
\newblock {\em Physical Review Letters}, 113(17), oct 2014.

\bibitem{Morin_2015}
F.~Morin, H.~Courtecuisse, M.~Chabanas, and Y.~Payan.
\newblock Rest shape computation for highly deformable model of brain.
\newblock {\em Computer Methods in Biomechanics and Biomedical Engineering},
  18(sup1):2006--2007, sep 2015.

\bibitem{murphy2003inverse}
J.~G. Murphy.
\newblock Inverse radial deformations and cavitation in finite compressible
  elasticity.
\newblock {\em Mathematics and Mechanics of Solids}, 8(6):639--650, 2003.

\bibitem{patte2022quasi}
C.~Patte, M.~Genet, and D.~Chapelle.
\newblock A quasi-static poromechanical model of the lungs.
\newblock {\em Biomechanics and Modeling in Mechanobiology}, 21(2):527--551,
  2022.

\bibitem{peirlinck2018modular}
M.~Peirlinck, M.~De~Beule, P.~Segers, and N.~Rebelo.
\newblock A modular inverse elastostatics approach to resolve the
  pressure-induced stress state for in vivo imaging based cardiovascular
  modeling.
\newblock {\em Journal of the mechanical behavior of biomedical materials},
  85:124--133, 2018.

\bibitem{pfaller2019importance}
M.~R. Pfaller, J.~M. H{\"o}rmann, M.~Weigl, A.~Nagler, R.~Chabiniok,
  C.~Bertoglio, and W.~A. Wall.
\newblock The importance of the pericardium for cardiac biomechanics: from
  physiology to computational modeling.
\newblock {\em Biomechanics and Modeling in Mechanobiology}, 18:503--529, 2019.

\bibitem{piersanti2021modeling}
R.~Piersanti, P.~C. Africa, M.~Fedele, C.~Vergara, L.~Ded{\`e}, A.~F. Corno,
  and A.~Quarteroni.
\newblock Modeling cardiac muscle fibers in ventricular and atrial
  electrophysiology simulations.
\newblock {\em Computer Methods in Applied Mechanics and Engineering},
  373:113468, 2021.

\bibitem{rathgeber2016firedrake}
F.~Rathgeber, D.~Ham, L.~Mitchell, M.~Lange, F.~Luporini, A.~McRae, G.-T.
  Bercea, G.~Markall, and P.~Kelly.
\newblock Firedrake: automating the finite element method by composing
  abstractions.
\newblock {\em ACM Transactions on Mathematical Software (TOMS)}, 43(3):1--27,
  2016.

\bibitem{rausch2017augmented}
M.~K. Rausch, M.~Genet, and J.~D. Humphrey.
\newblock An augmented iterative method for identifying a stress-free reference
  configuration in image-based biomechanical modeling.
\newblock {\em Journal of Biomechanics}, 58:227--231, 2017.

\bibitem{regazzoni2019reviewXB}
F.~Regazzoni, L.~Ded{\`e}, and A.~Quarteroni.
\newblock Active force generation in cardiac muscle cells: mathematical
  modeling and numerical simulation of the actin-myosin interaction.
\newblock {\em Vietnam Journal of Mathematics}, 49:87–118, 2021.

\bibitem{regazzoni2021oscillation}
F.~Regazzoni and A.~Quarteroni.
\newblock An oscillation-free fully partitioned scheme for the numerical
  modeling of cardiac active mechanics.
\newblock {\em Computer Methods in Applied Mechanics and Engineering},
  373:113506, 2021.

\bibitem{regazzoni2022emcirculation}
F.~Regazzoni, M.~Salvador, P.~C. Africa, M.~Fedele, L.~Dedè, and
  A.~Quarteroni.
\newblock A cardiac electromechanical model coupled with a lumped-parameter
  model for closed-loop blood circulation.
\newblock {\em Journal of Computational Physics}, 457:111083, 2022.

\bibitem{riccobelli2019existence}
D.~Riccobelli, A.~Agosti, and P.~Ciarletta.
\newblock On the existence of elastic minimizers for initially stressed
  materials.
\newblock {\em Philosophical Transactions of the Royal Society A: Mathematical,
  Physical and Engineering Sciences}, 377(2144):20180074, 2019.

\bibitem{Riccobelli_2019}
D.~Riccobelli and D.~Ambrosi.
\newblock Activation of a muscle as a mapping of stress–strain curves.
\newblock {\em Extreme Mechanics Letters}, 28:37–42, Apr. 2019.

\bibitem{Riccobelli_2017}
D.~Riccobelli and P.~Ciarletta.
\newblock {Rayleigh{\textendash}Taylor} instability in soft elastic layers.
\newblock {\em Philosophical Transactions of the Royal Society A: Mathematical,
  Physical and Engineering Sciences}, 375(2093):20160421, apr 2017.

\bibitem{rodriguez1994stress}
E.~K. Rodriguez, A.~Hoger, and A.~D. McCulloch.
\newblock Stress-dependent finite growth in soft elastic tissues.
\newblock {\em Journal of Biomechanics}, 27(4):455--467, 1994.

\bibitem{sellier2011iterative}
M.~Sellier.
\newblock An iterative method for the inverse elasto-static problem.
\newblock {\em Journal of Fluids and Structures}, 27(8):1461--1470, 2011.

\bibitem{Shield_1967}
R.~T. Shield.
\newblock Inverse deformation results in finite elasticity.
\newblock {\em Zeitschrift für Angewandte Mathematik und Physik {ZAMP}},
  18(4):490--500, jul 1967.

\bibitem{_ilhav__1997}
M.~{\v{S}}ilhav{\'{y}}.
\newblock {\em The Mechanics and Thermodynamics of Continuous Media}.
\newblock Springer Berlin Heidelberg, 1997.

\bibitem{taber2000modeling}
L.~A. Taber and R.~Perucchio.
\newblock Modeling heart development.
\newblock {\em Journal of Elasticity}, 61(1):165--198, 2000.

\bibitem{trilinos-website}
T.~{T}rilinos~{P}roject {T}eam.
\newblock {\em The {T}rilinos {P}roject {W}ebsite}, 2020 (acccessed May 22,
  2020).

\bibitem{truesdell2013non}
C.~Truesdell and W.~Noll.
\newblock {\em The Non-Linear Field Theories of Mechanics}.
\newblock Springer Science \& Business Media, 2013.

\bibitem{Usyk2002}
T.~P. Usyk, I.~J. LeGrice, and A.~D. McCulloch.
\newblock Computational model of three-dimensional cardiac electromechanics.
\newblock {\em Computing and Visualization in Science}, 4(4):249--257, 2002.

\bibitem{Zygote2014}
{Zygote}.
\newblock {Zygote solid 3D male anatomy collection generation II develompent
  report}.
\newblock Technical report, 2014.

\end{thebibliography}
\end{document}